\documentclass[aps,pre,showpacs,twocolumn]{revtex4-1}

\usepackage{amsmath,amssymb,amsfonts,hyperref}
\usepackage{graphicx,epsf}
\usepackage{xspace}
\usepackage{textcomp}
\usepackage{color}

\newif\ifhyper
\hypertrue
\ifhyper
\hypersetup{
  citecolor = {green},
  urlcolor = {blue} 
} 

\newcommand{\be}{\begin{equation}}
\newcommand{\ee}{\end{equation}}

\newcommand{\p}{\partial} 
\newcommand{\vx}{\vec{x}}

\newcommand{\bq}{{\bf q}}
\newcommand{\bp}{{\bf p}}
\newcommand{\bx}{{\bf x}}
\newcommand{\vq}{{\vec{q}}}
\newcommand{\vp}{{\vec{p}}}
\newcommand{\td}{\text{\tiny $D$}}
\newcommand{\xx}{\text{\tiny $X$}}
\newcommand{\xn}{\text{\tiny $0$}}

\newcommand{\tJ}{\tilde{j}}

\newcommand{\tih}{\tilde{h}}
\newcommand{\tj}{\tilde{j}}
\newcommand{\tom}{\hat{\omega}}
\newcommand{\tnu}{\hat{\varpi}}
\newcommand{\tp}{\hat{p}}
\newcommand{\tq}{\hat{q}}

\newcommand{\tg}{\hat{g}}
\newcommand{\tf}{\hat{f}}
\newcommand{\tI}{\hat{I}}
\newcommand{\baf}{\hat{\zeta}}
\newcommand{\fhat}{\tilde{f}}
\newcommand{\frond}{\mathring{f}}

\newcommand{\fy}{f}
\newcommand{\Fhat}{\tilde{F}}

\newcommand{\Frond}{\mathring{F}}
\newcommand{\Grond}{\mathring{H}}
 
\newcommand{\real}{{\rm Re}}
\newcommand{\ima}{{\rm Im}}

\newcommand{\etan}{\eta^{\nu}}
\newcommand{\etad}{\eta^{\td}}
\newcommand{\etax}{\eta^{\xx}}

\newcommand{\dis}{\displaystyle}

\newcommand{\npt}{nonperturbative\xspace}
\newcommand{\anz}{{ansatz}\xspace}

\newcommand{\tm}{\text{-}}

\begin{document}

\title{Nonperturbative renormalization group for the stationary Kardar-Parisi-Zhang equation: scaling functions and amplitude ratios in 1+1, 2+1 and 3+1  dimensions}

\author{Thomas Kloss$^1$, L\'eonie Canet$^1$, and  Nicol\'as Wschebor$^2$}
\affiliation{$^1$  Laboratoire de Physique et Mod\'elisation des Milieux Condens\'es, CNRS UMR 5493, Universit\'e Joseph Fourier Grenoble I, BP166,  38042 Grenoble Cedex, France\\$^2$Instituto de F\'isica, Facultad de Ingenier\'ia, Universidad de la Rep\'ublica, J.H.y Reissig 565, 11000 Montevideo, Uruguay}

\begin{abstract}

We investigate the strong-coupling regime of the stationary Kardar-Parisi-Zhang equation for interfaces growing on a substrate of dimension $d=1$, 2, and 3 using a nonperturbative renormalization group (NPRG) approach. We compute critical exponents, correlation and response functions, extract the related scaling functions and calculate universal amplitude ratios. We work with a simplified implementation of the  second-order (in the response field) approximation proposed in a previous work [PRE {\bf 84}, 061128 (2011) and Erratum {\bf 86}, 019904 (2012)],  which greatly simplifies the frequency sector of the NPRG flow equations, while keeping a nontrivial frequency dependence  for the 2-point functions. The one-dimensional scaling function obtained within this approach  compares very accurately  with the scaling function obtained from the full second-order NPRG equations and with the exact scaling function. 
 Furthermore, the approach is easily applicable to higher dimensions and we provide scaling functions and amplitude ratios in $d=2$ and $d=3$.
 We argue that our ansatz is reliable up to $d\simeq 3.5$.
\end{abstract}

\maketitle

\section{Introduction}

In their seminal paper \cite{kardar86}, Kardar, Parisi and Zhang (KPZ) proposed the nonlinear Langevin equation 
\begin{equation}
\frac{\p h(t,\vx)}{\p t} = \nu\,\nabla^2 h(t,\vx) \, + 
\,\frac{\lambda}{2}\,\big(\nabla h(t,\vx)\big)^2 \,+\,\eta(t,\vx)
\label{eqkpz}
\end{equation}
as a continuum model for the dynamics of a  growing interface. In this equation, $h(t,\vx)$ is a single valued height profile depending on the $d$-dimensional  substrate coordinate $\vx$ and on time $t$.
The term $\eta(t,\vx)$ represents an uncorrelated white noise with zero mean $\langle \eta(t,\vx)\rangle = 0$ and strength $D$, 
\begin{equation}
  \big\langle \eta(t,\vx)\eta(t',\vx')\big\rangle = 2\,D\,\delta^d(\vx-\vx')\,\delta(t-t'), 
\end{equation}
which accounts for the randomness in the growing mechanism. The term  $\nu\nabla^2 h$ describes the surface tension which tends to smoothen the interface. The nonlinear term, which represents an excess growth along the local normal to the surface, is the essential ingredient to capture  the nontrivial behavior of the growing interface. 
The KPZ equation generates generic scaling
 (without fine-tuning any parameter). For instance, the 2-point, height-height correlation function in the co-moving frame
\begin{equation}
C(t,\vx) = \langle h(t,\vx) h(0,0) \rangle_c
\label{eq:corrfunc}
\end{equation}
endows at large time $t$  and distance $x = |\vx|$, the dynamical scaling form 
\begin{equation}
  C(t,\vx) = x^{2\chi}\,F(t / x^z), 
\label{eq:corrfunc2}
\end{equation}
where $\chi$ and $z$ are the universal  roughness and dynamical critical exponents and  $F(y)$ is a universal scaling function. The KPZ equation (\ref{eqkpz}) encompasses a nonequilibrium phase transition in dimension $d>2$ between a smooth ($\chi = 0$) and a rough ($\chi > 0$)  phase, separated by a critical value $\lambda_c$ of the non-linearity. In  dimension $d \leq 2$, the interface always roughens.

In fact, the KPZ equation arises in connection with a large class of nonequilibrium and disordered systems and has thus emerged as one of the fundamental theoretical models for nonequilibrium phase transitions and scaling phenomena \cite{halpin-healy95,krug97}. Despite its apparent simplicity,
the KPZ equation has resisted most of the theoretical attempts to provide a complete description of its strong coupling phase in generic dimensions. Very recently, an impressive breakthrough has been achieved to describe one-dimensional interfaces, bringing out both exact analytical solutions  \cite{Johansson00,praehofer04,sasamoto05,*sasamoto10a,*sasamoto10b,calabrese11,Amir11,Imamura12,Corwin12} and high-precision experimental measurements \cite{takeuchi10,*takeuchi11,*Takeuchi12}. An unanticipated connection with random matrix theory has been revealed
 theoretically and neatly confirmed experimentally. Namely, the height fluctuations were shown to follow a Tracy-Widom distribution, of either the Gaussian unitary ensemble (GUE) or the Gaussian othogonal ensemble (GOE) depending on the geometry of the interface, respectively  curved or flat \cite{Johansson00,praehofer04,sasamoto05,*sasamoto10a,*sasamoto10b,calabrese11,Amir11,Imamura12,Corwin12}. Note that a generalization of this ideas to 2+1 dimensions has been recently explored by means of extensive Monte Carlo simulations \cite{Halpin-Healy12}.

From the theoretical point of view, the one-dimensional problem is special and much simpler due to the existence of an additional  generalized fluctuation-dissipation theorem (FDT) in this dimension.  Thereby, the exponents are known exactly as $z = 3/2$ and $\chi = 1/2$.
However, in dimension $d>1$, the KPZ equation remains  unsolved.
 This deceptive situation has persisted due to a lack of controlled analytical tools to describe the strong-coupling regime. 
 Notably,  standard perturbative expansions and renormalization group (RG) methods \cite{kardar86,medina89,nattermann92,frey94}  were proved to fail {\it at all orders} to find a strong coupling fixed point \cite{wiese98}. Some nonperturbative approaches have been devised, such as the mode-coupling (MC) approximation  
\cite{beijeren85,frey96,bouchaud93,colaiori01a}, 
  the self-consistent expansion (SCE) \cite{schwartz92,*schwartz08},
 or the weak noise scheme \cite{fogedby01,fogedby05,*fogedby06}. Beside critical exponents, they provided some results concerning the one-dimensional scaling functions and various (conflicting) predictions regarding the existence of an upper 
critical dimension, above which $\chi = 0$,  that is the interface would remain smooth for all values of the nonlinearity.

In the present work, we  concentrate on the universal scaling functions and amplitude ratios related to  the KPZ equation.
The existence in generic dimensions of  a scaling form for the stationary correlation function of the KPZ growth and of universal amplitude ratios was early observed  \cite{hwa91,amar92R,amar92}. 
This observation was corroborated in $d=1$ with reasonable agreement by many different approaches, 
including Monte-Carlo simulations of discrete models, direct numerical integration of the MC equations \cite{tang92,krug92,frey96} and a soliton approximation \cite{fogedby01}.
In subsequent studies, a stretched exponential decay of the tail of the scaling function  was predicted independently by the SCE \cite{schwartz02} and the MC \cite{colaiori01a,colaiori01b} approximations, and numerically verified in one dimension within both approaches \cite{Colaiori01c,katzav04}. 
Later, the one-dimensional scaling function was calculated exactly in the long-time limit \cite{praehofer04} (and for all the scaling regime very recently \cite{Imamura12}).
The exact stationary scaling function in $d = 1$ shows an exponential tail, but on a different scale than that brought out  by MCT and SCE approximations \cite{colaiori01a,colaiori01b,katzav04}.
On the other hand, in higher dimensions, apart from the prediction of the asymptotics of the tail from  SCE and MCT, theoretical calculations of scaling functions or amplitude ratios  are scarce, if not inexistent \footnote{To our knowledge, only Ref.\ \cite{Tu94} reports a two-dimensional calculation of the scaling functions, relying on a numerical integration of the MC equations. However, the obtained values for the critical exponents lead to the conclusion of an absence of upper critical dimension, which was later invalidated by more accurate MC calculations \cite{colaiori01a}.}. 
 Only some numerical simulations are available in higher dimensions, and have provided the width distribution   \cite{marinari02,kelling11}.

Over the last decade, the nonperturbative renormalization group (NPRG) has emerged as a powerful tool to investigate strongly correlated systems and critical phenomena  \cite{berges02,*Delamotte07,*Kopietz10}. 
Besides a widespread application for equilibrium problems, the NPRG was further extended to study stationary \cite{canet04a,canet11b} and also transient nonequilibrium systems (see {\it e.g}.\ \cite{Kloss11,Berges12}).
In Ref.\ \cite{canet10},  the NPRG techniques were first applied 
 to study the stationary KPZ equation in $d$ dimensions for an overall flat geometry. A simple ansatz was devised,  referred to  as leading-order (LO) in the following. 
 This simple approximation already entails generic scaling and  correctly captures the strong-coupling behavior
 in all dimensions, providing  critical exponents at the strong coupling fixed point \cite{canet10} which are consistent with existing results in physical dimensions.

A  general and systematic framework to apply NPRG techniques to the KPZ equation is presented in Ref. \cite{canet11a}. In this work, the symmetries of the KPZ equation are analyzed in detail and the related Ward identities are derived. A `covariantization' associated with the Galilean invariance is introduced,  which provides a simple geometrical interpretation of this symmetry, useful to formulate ansaetze respecting it. 
An appropriate NPRG approximation scheme,  strongly constrained by all these symmetries, is devised. The second-order (SO) approximation (corresponding to a truncation at quadratic order in the response field), is explicitly implemented and  the SO flow equations are solved in $d=1$ to obtain the scaling functions and the  amplitude ratio for the KPZ equation in this dimension, again for a flat geometry \cite{canet11a}. 
The scaling functions turn out to compare very accurately with the exact ones \cite{praehofer04}, reproducing in particular all the fine structure of the tail on the correct scale.
However, the one-dimensional case remains special due to the incidental existence of the generalized FDT and the  NPRG flow equations in $d=1$ are simpler. The numerical solution of the SO ansatz in arbitrary dimensions is  more involved.

In the present work, we propose a simplification of the SO ansatz which allows us to solve the corresponding flow equations in arbitrary dimensions without cumbersome numerics. We will refer to this approximation as next-to-leading order (NLO). The simplification consists in neglecting the  frequency dependence of some flowing functions in the integrals of the NPRG  equations (see Sec. \ref{ANZ}). This approximation drastically simplifies the frequency sector of the  flow equations, enabling one to calculate analytically all the frequency integrals.
It results in a reduction of  the computing time for the numerical integration of the flow equations by a factor of around 200, which in turn  allows us 
 to achieve an extensive study of the NLO flow equations in various dimensions.

The physical quantities of interest in this paper are the correlation and response functions of the stationary KPZ equation in the strong coupling regime (and for a flat geometry).
From the correlation and response functions we extract the related scaling functions and determine universal amplitude ratios.
We first study the one-dimensional case. The scaling functions obtained at NLO, although less precise than with the complete SO approximation, still compare accurately with their exact counterparts. We then investigate
 higher dimensions. We provide the  scaling functions and universal amplitude ratios in   $d = $ 2 and 3. To the best of our knowledge, no other comparable results are available in the literature. The obtained scaling functions show an exponential decay, but on a different scale than that predicted by SCE and MCT \cite{colaiori01a,schwartz02}, as in $d=1$. We also determine the deviations in higher dimensions from the generalized FDT, which is only exactly fulfilled in $d = $ 1. 

The validity of  the NLO approximation can be assessed in two respects:   in $d=1$ by confronting the scaling functions with the exact results as mentioned before; in $d>1$ through a detailed analysis of the critical exponents. First we find at NLO values in  $d = $ 2 and 3 that are consistent with  results from numerical works reported in the literature \cite{tang92,marinari00}. Moreover, we investigate the dependence of our results on the NPRG regulator, which provides us with some further check of the quality of the approximation. Although the NLO approximation deteriorates with the dimension,  we argue that our NLO ansatz is reliable up to $d \simeq $ 3.5. The NLO ansatz is therefore  applicable to study the KPZ equation in dimension 2 and 3,  but cannot  contribute to probe the existence of an upper critical dimension.

The remainder of the paper is organized as follows. 
In Section \ref{NPRG}, we briefly review the \npt renormalization group formalism for out-of-equilibrium problems. Section \ref{ANZ}  presents the symmetries and a general NPRG approximation scheme  for the KPZ problem. We also specify the NLO approximation and  derive the corresponding flow equations.
 The calculation of the fixed point properties is presented in Section \ref{RES}. We provide   critical exponents, amplitude ratios and  universal scaling functions.  Section \ref{conclusion} contains  a summary and concluding remarks. Technical details are postponed in the Appendices.

\section{The \npt renormalization group}
\label{NPRG}

The general NPRG formalism for nonequilibrium systems is expounded in details in Ref.\  \cite{canet11b,Berges12}. Its application to the KPZ equation is further specified  in Ref.\  \cite{canet11a}. Consequently, we recall  for completeness only the main elements of this formalism, essentially following Ref.\ \cite{canet11a}.  In particular, the Fourier conventions used throughout this work are:
\begin{subequations}
\begin{align}
 f(\omega,\vp)  \! &= \! \! \! \int \! \! d^d \vec x \,dt \! \;f(t,\vx) \, e^{-i \vp \cdot \vx + i\omega t} \! \!
\equiv  \! \!\int_{\bf x} \! \! \! f({\bf x})\! \; e^{-i \vp \cdot\vx + i\omega t} , \\
 f(t,\vx) \! &=  \! \! \! \int \! \! \frac{d^d \vp}{(2\pi)^{d}} \frac{d \omega}{2\pi} \,  f(\omega,\vp)\, 
e^{i \vp \cdot \vx - i\omega t} 
\!\! \equiv \!\! \int_{\bp} \! \! \! f (\bp)\, e^{i \vp \cdot\vx - i\omega t},
\end{align}
\end{subequations}
where ${\bf x}=(t,\vx)$ and $\bp=(\omega,\vp)$ and the factor $(2\pi)^{d+1}$ is implicitly included in the momentum integral.

The starting point is the field theory associated with
 Eq.\  (\ref{eqkpz}), which can be derived from the standard Janssen-de Dominicis procedure \cite{janssen76, *dominicis76} relying on the introduction of a Martin-Siggia-Rose \cite{Martin73} response field  $\tih$ and sources $(j,\tj)$ . 
The generating functional reads:
\begin{subequations}
\begin{align}
\!\!\!\!{\cal Z}[j,\tj] \! &= \!\!\int \!{\cal D}[h,i \tih]\, 
\exp \! \left(-{\cal S}[h,\tih] +  \int_{\bf x} \left(j h+\tj\tih\right) \right) , \label{Z}\\
\!\!\!\!{\cal S}[h,\tih]  \! &= \!\! \int_{\bf x}  \!\left\{ \tih\left(\p_t h -\nu \,\nabla^2 h - 
\frac{\lambda}{2}\,({\nabla} h)^2 \right) - D\, \tih^2  \right\} .
\label{ftkpz}
\end{align} 
\end{subequations}
Averages of an observable ${\cal O}$ over the disorder are expressed in terms of the functional integral
\begin{align}
\langle {\cal O}  \rangle
= \frac{\int \!{\cal D}[h,i \tih]\,  {\cal O}\,e^{-{\cal S}[h,\tih]}}{\int \!{\cal D}[h,i \tih]\, e^{-{\cal S}[h,\tih]}  } .
\label{funcAver}
\end{align} 

The NPRG formalism  consists in building  a sequence of scale-dependent  effective models   
such that fluctuations are smoothly averaged as the (momentum) scale $\kappa$ is lowered 
from the  microscopic scale $\Lambda$, where no fluctuations are yet included, to the macroscopic scale $\kappa=0$, where  all fluctuations 
 are summed over \cite{berges02,delamotte05}.
For out-of-equilibrium problems, one formally proceeds as in equilibrium,
 but with the presence of the additional response fields and with special care required to deal with  the consequences of It$\bar{\rm o}$'s discretization and with causality issues,
 as stressed in detail in \cite{canet11b,Bernitez12b}. 

To achieve the separation of fluctuation modes within the NPRG procedure, one  adds to the original action ${\cal S}$
a momentum and scale dependent mass-like term:
\begin{equation}
\Delta {\cal S}_\kappa \!=\!\frac{1}{2}\! \int_{\bf q}\!  h_i(-{\bf q})\, 
[R_\kappa({\bf q})]_{ij}\, h_j({\bf q}) ,\;\;\label{deltask}
\end{equation}
where the indices $i,j\in\{1,2\}$ label the field and response field, respectively $h_1=h,h_2=\tih$, and 
summation over repeated indices is implicit. 
The matrix elements $[R_\kappa({\bf q})]_{ij}$ are proportional to a cutoff function
 $r(q^2/\kappa^2)$, with $q=|\vq|$, which ensures the selection of 
fluctuation modes: $r(x)$ is required to vanish as $x\gtrsim 1 $ such that 
the fluctuation modes $h_i(q \gtrsim \kappa)$ are unaffected by 
$\Delta {\cal S}_\kappa$,
and to be large  when $x\lesssim 1 $ such that the other modes ($h_i(q\lesssim \kappa)$) are essentially frozen. Furthermore,  $\Delta {\cal S}_\kappa$ must preserve all the symmetries of the problem and causality properties. An appropriate choice is 
\begin{equation}
R_\kappa(\omega,\vq) \!\equiv\!R_\kappa(\vq) \!=\! r\left(\frac{q^2}{\kappa^2}\right)
\left(\!\! \begin{array}{cc}
0& {\nu_\kappa} q^2\\
{\nu_\kappa} q^2 & -2 D_\kappa 
\end{array}\!\!\right) \;,
\label{Rk}
\end{equation}
where the running coefficients $\nu_\kappa$ and $D_\kappa$, defined later (Eq.\ (\ref{eq:dknukdef})),  are introduced in the regulator for  convenience \cite{canet10}. 
Here we choose the cutoff function to be 
\begin{equation}
r(x)=\alpha/(\exp(x) -1), 
\label{eq:expReg}
\end{equation}
where $\alpha$ is a free parameter.

In presence of the mass term $\Delta {\cal S}_\kappa$, the generating functional (\ref{Z}) becomes scale dependent 
\begin{equation}
{\cal Z}_\kappa[j,\tj] \!\! = \!\!\!\int {\cal D}[h,i \tih]\, 
\exp\left(-{\cal S}-\Delta{\cal S}_\kappa+  \int_{\bf x} \left( j h+\tj\tih \right)\right) . \label{Zk}
\end{equation} 
  Field expectation values in the presence of the external sources $j$ and $\tilde{j}$ are obtained from the functional  ${\cal W}_\kappa = \log {\cal Z}_\kappa$ as 
\begin{equation}
  \varphi({\bf x}) = \langle h({\bf x}) \rangle = \frac{\delta {\cal W}_{\kappa}}{\delta j({\bf x})}  \, \, , \, \,
  \tilde \varphi({\bf x}) = \langle \tilde h({\bf x}) \rangle = \frac{\delta {\cal W}_{\kappa}}{\delta \tilde j({\bf x})}  .
\end{equation} 
The effective action $\Gamma_\kappa[\varphi,\tilde\varphi]$ is now given by the Legendre transform of  ${\cal W}_\kappa$
(up to a term proportional to $R_\kappa$) \cite{berges02,*Delamotte07,*Kopietz10,canet11b}:
\begin{equation}
\Gamma_\kappa[\varphi,\tilde\varphi] +{\cal W}_\kappa[j,\tj] = 
\int \! j_i \varphi_i -\frac{1}{2} \int \varphi_i \, [R_\kappa  ]_{ij}\, \varphi_{j} .
\label{legendre}
\end{equation}
 From $\Gamma_\kappa$, one can derive 2-point functions
\be
[\,\Gamma_\kappa^{(2)}\,]_{i_1 i_2}({\bf x}_1,{\bf x}_2, \varphi,\tilde\varphi) = 
\frac{\delta^2 \Gamma_\kappa[\varphi,\tilde\varphi]}{\delta\varphi_{i_1}({\bf x}_1)\delta\varphi_{i_2}({\bf x}_2)} , 
\ee
and more generally $n$-point 1-PI (one particle irreducible) functions, that we write 
 here
in a $2\times2$ matrix form  (omitting the dependence on the running scale $\kappa$ when $n>2$)
\begin{equation}
\Gamma_{i_3,...,i_n}^{(n)}({\bf x}_1,...,{\bf x}_n,\varphi,\tilde\varphi) =
\frac{\delta^{n-2} \Gamma_\kappa^{(2)}({\bf x}_1,{\bf x}_2,\varphi,\tilde\varphi)}
{\delta\varphi_{i_3}({\bf x}_3)...\delta\varphi_{i_n}({\bf x}_n)} \;.
\end{equation}
The matrix conventions are taken from Ref.\  \cite{canet11b}.
The exact flow for $\Gamma_{\kappa}[\varphi,\tilde\varphi]$ is given by Wetterich's 
equation, which  is in Fourier space  \cite{berges02,*Delamotte07,*Kopietz10}
\begin{equation}
\partial_\kappa \Gamma_\kappa = \frac{1}{2}\, {\rm Tr}\! \int_{\bf q}\! \partial_\kappa R_\kappa \cdot G_\kappa ,
\label{dkgam}
\end{equation}
where
\begin{equation}
 G_\kappa=\left[\Gamma_\kappa^{(2)}+R_\kappa\right]^{-1}
 \label{eq:propag}
\end{equation}
is the full renormalized propagator of the theory.
When $\kappa$ flows from $\Lambda$ to 0, $\Gamma_\kappa$ interpolates 
between the microscopic model $\Gamma_{\kappa=\Lambda}={\cal S}$  and the full effective action $\Gamma_{\kappa=0}$
 that encompasses all the macroscopic properties of the system \cite{canet11b}.
Differentiating Eq.\  (\ref{dkgam}) twice with respect to the fields
and evaluating it in a uniform and stationary field configuration (since the model is analyzed in its long time and large distance regime where it is translationally invariant in space and time)
one obtains the flow equation for the 2-point functions
\begin{eqnarray}
\partial_\kappa [\,\Gamma^{(2)}_\kappa\,]_{ij}(\bp)\! &=& \! {\rm Tr}\! \int_{\bq} \partial_\kappa R(\bq) \cdot G_\kappa(\bq) \cdot
\!\bigg(\!\!-\!\frac{1}{2}\, \Gamma^{(4)}_{ij}(\bp,-\bp,\bq) \nonumber\\
&& \hspace{-2.2cm} +\Gamma^{(3)}_{i}(\bp,\bq) \cdot G_\kappa(\bp+\bq) \cdot
\Gamma^{(3)}_{j}(-\bp,\bp+\bq) \bigg) \cdot G_\kappa(\bq) ,
\label{dkgam2}
\end{eqnarray}
where the  background field dependences have been omitted,
as well as the last  argument of the $\Gamma^{(n)}$, which is determined by
frequency and momentum conservation \cite{canet11b}.\\

Solving Eq.\  (\ref{dkgam}) (or Eq.\  (\ref{dkgam2})) is in principle equivalent 
to solving the model. In practice this resolution cannot be performed exactly since these equations  are  nonlinear integral partial differential functional equations. Hence one has to devise an approximation scheme, adapted to the specific model under study and in particular to its symmetries.

\section{Approximation scheme}
\label{ANZ}

\subsection{Symmetries}
\label{sec:symm}
The approximation scheme must preserve the symmetries of the problem.
 The KPZ action (\ref{ftkpz}) possesses three  symmetries, besides the  translational and rotational invariances: i) the Galilean invariance, which translates for an interface as the invariance under an infinitesimal tilting,  ii) the invariance under a constant shift of the $h$ field, and iii) an additional time reversal  symmetry only realized in $d=1$. These symmetries can be expressed as an invariance of the action (\ref{ftkpz}) under the following field transformations
\begin{eqnarray}
\text{i)} && \left\{
\begin{array}{l}
h'(t,\vec x)=\vec x \cdot \vec v + h(t,\vec x+ \lambda \vec v t)\\
\tilde h'(t,\vec x)=\tilde h(t,\vec x+ \lambda \vec v t) \label{gal}
\end{array}
\right.\\
\text{ii)} && \;\;\;\; h'(t,\vec x)=h(t,\vec x)+c, \label{shift}
\end{eqnarray}
where $\vec v$ and $c$ are arbitrary constant quantities. 
As for the time reversal symmetry, it can be encoded in the field transformation
 \cite{canet05,frey96}
\begin{equation}
\text{iii)} \left\{
\begin{array}{l}
h'(t,\vx)=- h(-t,\vx) \\
\tilde h'(t,\vx)=\tilde h(-t,\vx) +\frac{\nu}{D} \nabla^2 h(-t,\vx)
\end{array}
\right.  .\label{fdt}
\end{equation}
 One can check that the term $\int_{\bf x}  \tilde h \p_t h$ and the combination $\int_{\bf x} ( \nu\tilde h \nabla^2 h +D\tilde h^2)$ of the action (\ref{ftkpz}) are both invariant under the transformations (\ref{fdt}) in any dimensions. On the other hand, the variation of the nonlinear term is proportional to $\int_{\bf x} (\nabla h)^2 \nabla^2 h$, which is non vanishing
in generic dimensions. However, in $d=1$, it becomes the integral of a total derivative $\int_{\bf x}\partial_x(\partial_x h)^3$ that vanishes provided  $\partial_x h$  tends to zero at large  $x$, resulting in the existence of a time-reversal symmetry only in this dimension \footnote{Note that, as tested and discussed in  Appendix D, the NLO approximation presented below remains very accurate, even if the time-reversal symmetry is not imposed.}.

These symmetries are analyzed in details in  \cite{frey96,canet11a}, where the corresponding Ward identities are derived.
 In fact, the symmetries i) and ii) admits a time-gauged form \cite{canet11a,lebedev94},  expressed as  the following infinitesimal field transformations:
 \begin{eqnarray}
\text{i')} && \left\{
\begin{array}{l}
h'(t,\vx)=\vx \cdot \p_t \vec v(t) + h(t,\vx+ \lambda \vec v(t))\label{galg}\\
\tilde h'(t,\vx)=\tilde h(t,\vx+ \lambda \vec v(t))
\end{array}
\right.\\
\text{ii')} && \;\;\;\; h'(t,\vx)=h(t,\vx)+c(t). \label{timeg}
\end{eqnarray}
As the  variations of the KPZ action under these time-gauged transformations are linear in the fields, they  also entail simple Ward identities, which extend with a stronger contents the ones deduced from the non-gauged forms \cite{canet11a}.

Regarding the shift-gauged symmetry, we  simply recall a useful property of the $n$-point vertex functions, denoted from now on  as 
\begin{equation}
\Gamma_\kappa^{(l,m)}({\bf x}_1,\dots,{\bf x}_{l+m}),
\end{equation}
which stands for  the $\Gamma_\kappa^{(n=l+m)}$ vertex involving $l$ (respectively  $m$)
legs -- derivatives of $\Gamma_\kappa$ with respect to $\varphi$ (respectively  $\tilde\varphi$) -- with the $l$ first frequencies and 
momenta referring to the $\varphi$ fields and the $m$ last to the $\tilde\varphi$ fields. 
The shift-gauged symmetry entails that the $n$-point vertex functions in Fourier space satisfy the following property:
\begin{equation}
\label{wardtemp}
\Gamma_\kappa^{(m,n)}(\omega_1,\vec p_1=0,\dots,{\bf p}_{m+n-1})= i\omega_1 \delta_{m1}\delta_{n1}
\end{equation}
 for any $m\geq 1$. This means that, apart from the contribution of  $\int \tilde \varphi \partial_t \varphi$  to  $\Gamma_\kappa^{(1,1)}$, the vertices vanish upon setting  the momentum of one of the  $\varphi$ to zero.

As for the time-reversal symmetry, we just give the Ward identity which relates the 2-point vertex functions in $d=1$
\begin{equation}
2\Re\Gamma_\kappa^{(1,1)}({\bf p})
=-\frac{\nu}{D}p^2\Gamma_\kappa^{(0,2)}({\bf p}),
\label{wardfdt}
\end{equation}
which will be useful in the following.

Finally,  the Ward identities ensuing from the gauged Galilean symmetry, relating the momentum and frequency sectors of vertex functions with different number of legs bare  complicated forms and can be found in Ref. \cite{canet11a}.
 Let us stress instead a simple geometrical interpretation of this symmetry proposed in \cite{canet11a}. A function $f(\bx)$ is defined as a scalar under the Galilean transformation (\ref{galg}) if its infinitesimal variation under this transformation is $\delta f(\bx) =\lambda \vec v(t)\cdot\nabla f(\bx)$. This then implies that $\int d^d\vec x f(\bx)$ is invariant under a Galilean transformation. With this definition $\tilde h$ and  $\nabla_i\nabla_j h$ are scalars, but  $\p_t h$ is not.  However, a scalar can be built from it if one introduces  a covariant form of the time derivative $D_t h$, defined as
\begin{equation}
 D_t h({\bf x})\equiv \partial_t h({\bf x})-\frac{\lambda}{2} (\nabla h({\bf x}))^2.
\end{equation}
 These three basic scalars  can  be combined together through sums, products and  spatial gradients $\nabla$  to construct an action manifestly invariant under the Galilean transformation. The time derivative of a scalar is not a scalar itself, but one can again build a scalar quantity by defining a general covariant time derivative:
\begin{equation}
\tilde{D}_t = \partial_t -\lambda \nabla h({\bf x})\cdot \nabla ,
\label{covardt1}
\end{equation}
which preserves the scalar property. 
 This construction will ground the derivation of our \anz.

\subsection{Rescaling}
\label{sec:Rescaling}

Before we proceed, note that the bare parameters $\nu$ and $D$ of the original KPZ action (\ref{ftkpz}) can be absorbed upon rescaling the fields and time
as
\be
 h\to\sqrt{\frac{D}{\nu}} h \quad,\quad  \tih\to \sqrt{\frac{\nu}{D}}\tih \quad,\quad  t\to t/ \nu .
\label{rescale}
\ee
 Followingly, the rescaled theory depends on a single effective coupling constant 
\begin{equation}
   \lambda \to \left( \frac{\lambda^2 D}{\nu^{3}} \right)^{1/2} \equiv \sqrt{g_b} 
   \label{eq:lambdaRes}
\end{equation} 
and the rescaled KPZ action reads
\begin{align}
\!\!\!\!{\cal S}[h,\tih]  \! &= \!\! \int_{\bf x}  \!\left\{ \tih\left(\p_t h -\nabla^2 h - 
\frac{\sqrt{g_b}}{2}\,({\nabla} h)^2 \right) - \tih^2  \right\} 
\label{ftkpzres}
\end{align} 
which amounts to setting $D$ and $\nu$ to unity and changing $\lambda$ to $\sqrt{g_b}$ in the original theory.
Accordingly, the covariant time derivatives Eq.\ (\ref{covardt1})  in the rescaled theory  read
\begin{subequations}
\begin{align}
D_t h({\bf x}) &=  \partial_t  h({\bf x})  -\frac{\sqrt{g_b}}{2} (\nabla h({\bf x}))^2  , \\
\tilde{D}_t &=  \partial_t -\sqrt{g_b} \nabla h({\bf x})\cdot \nabla  .
\end{align}
\label{covardt2}
\end{subequations}
For future use, we give the relation  between the original (l.h.s.) and the rescaled  (r.h.s.) $n$-point vertex functions in Fourier space, which is
\begin{align}
&\Gamma_\kappa^{(l,m)}({\bf p}_1,\dots,{\bf p}_{l+m-1}) \big|_{\text{original}}  \nonumber \\
& \qquad = \left( \frac{D}{\nu}\right)^{\frac{m-l}{2}}\nu \,\Gamma_\kappa^{(l,m)}({\bf p}_1',\dots,{\bf p}_{l+m-1}') \big|_{\text{rescaled}} 
\end{align}
with $ {\bf p}' = (\omega/\nu,\vp)$.
Unless stated otherwise, we work from now on with the rescaled theory.

\subsection{NLO ansatz}
\label{sec:NLOansatz}

Our aim is to compute  the physical correlation and response functions in the scaling regime. Consequently, 
 our ansatz is designed to provide an accurate description of the 2-point vertex functions, with  arbitrary  momentum and frequency dependences, while approximating higher order vertices. To do so, one can resort in general in the NPRG framework to 
 the Blaizot-M\'endez-Wschebor (BMW) 
approximation scheme, which  is precisely devised to compute momentum dependent $n$-point functions and has been proved to yield very accurate
results	for the	$O(N)$ model \cite{blaizot06,*benitez09,*benitez09}. However,  the symmetries of the KPZ equation
prevent  from a direct application of the BMW scheme.
 Nevertheless,   in Ref.\ \cite{canet11a},  the BMW strategy was adapted to cope with the KPZ symmetries and an approximation scheme for the KPZ problem was built. An explicit \anz for $\Gamma_\kappa$   was proposed, which we call second order (SO). It is truncated at quadratic order in the response field $\tilde h$, while preserving the complete momentum and frequency dependence of the 2-point functions.
The SO \anz   reads
\begin{align}
\Gamma_\kappa[\varphi,\tilde \varphi]=&   \dis \int_{\bf x} 
\left\{ \tilde \varphi f_\kappa^\lambda(\tm\tilde D_t^2,\tm\nabla^2) D_t\varphi - 
\tilde \varphi f_\kappa^\td(\tm\tilde D_t^2,\tm\nabla^2) \tilde \varphi \right. \nonumber \\
  &\hspace*{-10ex} -\frac 1 2 \left[\nabla^2 \varphi f_\kappa^\nu(\tm\tilde D_t^2,\tm\nabla^2) \tilde \varphi + \left. \tilde \varphi f_\kappa^\nu(\tm\tilde D_t^2,\tm\nabla^2)  \nabla^2 \varphi\right] \right\},
\label{anznlo}
\end{align}
where the three running functions $f_\kappa^\xx$, $X\in \{ \lambda, \nu,D \}$, are assumed to admit a formal expansion in a series of their arguments 
\begin{equation}
\label{serieformelle}
 f_\kappa^\xx(-\tilde D_t^2,-\nabla^2)=\sum_{m,n=0}^\infty a_{mn} (-\tilde D_t^2)^m (-\nabla^2)^n ,
\end{equation}
that is, all the $\tilde D_t$ operators act on the left of the $\nabla$ ones.
At the microscopic scale  $\kappa=\Lambda$, the three functions are constants 
$f_\Lambda^\xx \equiv 1$, and the (rescaled) KPZ action (\ref{ftkpzres}) is recovered.

\subsubsection{Running anomalous dimensions}

We introduce two scale dependent parameters $D_\kappa$ and $\nu_\kappa$ defined as
\begin{equation}
 D_\kappa \equiv f_\kappa^\td(0,0)\quad , \quad
\nu_\kappa \equiv f_\kappa^\nu(0,0)
\label{eq:dknukdef}
\end{equation}
to allow the fields and the dynamical scaling of time to  renormalize during the flow. In the rescaled theory, both parameters are dimensionless and at the bare scale, $D_\Lambda = \nu_\Lambda =1$. 
On the other hand, the function $f_\kappa^\lambda$ is fixed to unity at vanishing arguments  $f_\kappa^\lambda(0,0) = 1$ for all $\kappa$ by the shift-gauged symmetry (see Eq.\ \ref{eq:shiftgauge}). 
We define two anomalous dimensions associated with the two  running coefficients as
 \begin{equation}  
 \etad_\kappa = -\kappa \partial_\kappa \ln D_\kappa \quad , \quad 
 \etan_\kappa = -\kappa\partial_\kappa\ln \nu_\kappa .
  \label{eq:etaflow}  
 \end{equation}   
Their connection with the physical critical exponents is specified in Sec. \ref{RES}.

\subsubsection{Running vertex functions}

The calculation of the 2-point functions with the \anz (\ref{anznlo}) is straightforward and yields
\begin{subequations}
\begin{eqnarray}
\Gamma_\kappa^{(2,0)}(\omega,\vp) &=& 0 , \\
\Gamma_\kappa^{(1,1)}(\omega,\vp) &=& i \omega \,f_\kappa^\lambda\left(\omega, p\right) +  \vp\,^2 f_\kappa^\nu(\omega,p)  , \\
\Gamma_\kappa^{(0,2)}(\omega,\vp) &=& -2  f_\kappa^\td(\omega,p)  ,
\end{eqnarray}
\label{anzgam2}
\end{subequations}
where $p=|\vp|$, and noting that the actual dependence of the three running functions $f_\kappa^\xx$ is on $\omega^2$ and $p^2$.
 The  shift-gauged Ward identity Eq.\  (\ref{wardtemp}) further imposes the identity 
\begin{equation}
 f_\kappa^\lambda(\omega,p=0)=1 \quad\quad \hbox{for all $\kappa$}.
 \label{eq:shiftgauge}
\end{equation}
The expressions of the 3- and 4-point functions are lengthy and reported in  Ref.\  \cite{canet11a}.
One can check that they fulfill the Ward identities related to the three symmetries from Sec.\ \ref{sec:symm}, see Ref.\  \cite{canet11a}.

Note that $d=1$ is a special case  since the additional time-reversal symmetry
 leads to further constraints for the running functions.
 First the  Ward identity (\ref{wardfdt}) imposes 
\begin{equation}
f^\nu_\kappa(\omega, p)=f^\td_\kappa(\omega, p) \equiv f_\kappa(\omega, p) \quad, \quad D_\kappa=\nu_\kappa
\label{eq:d1f} 
\end{equation}
 and followingly
\begin{equation}
\etan_\kappa=\etad_\kappa  \equiv \eta_\kappa.
\label{eq:d1eta} 
\end{equation}
 Next, to consistently preserve this identity within the SO approximation, the entire $f^\lambda_\kappa(\omega,p)$ function has to be fixed to one in this dimension (see Appendix C)
\begin{equation}
f^\lambda_\kappa(\omega, p)\equiv 1,
\label{eq:d1flam} 
\end{equation}
such that  one is left in $d=1$ with a single independent function $f_\kappa$ and a single anomalous dimension $\eta_\kappa$ \cite{canet11a}.

In this paper, we work with the SO ansatz but implementing an additional approximation, which we refer to  as next-to-leading order (NLO). The motivation is that, while all the $\Gamma_\kappa^{(n,m)}$ functions and thus the flow equations greatly simplify in one dimension thanks to the additional time reversal symmetry, they remain complicated in generic dimensions. 
The NLO approximation
is performed in the flow equations for the 2-point functions given by Eq.\  (\ref{dkgam2}).
It consists in neglecting in the integrals on the right-hand side (r.h.s.)\ of the flow equation (\ref{dkgam2}) the frequency dependence of the three running functions  $f_\kappa^\xx$, hence only keeping in the integrands  the explicit frequency dependence of $\Gamma_\kappa^{(1,1)}$. 
The NLO approximation thus consists in performing the replacement
\begin{equation}
 f_\kappa^\xx(\omega,p)  \rightarrow f_\kappa^\xx(p)
 \label{eq:nloreplace}
\end{equation}
 in all $n$-point vertex functions on the r.h.s.\ of the flow equation (\ref{dkgam2}), and in the propagator $G_\kappa$.
The substitution Eq.\  (\ref{eq:nloreplace}) in the 3- and 4-point vertex functions yields that they all vanish except  $\Gamma_\kappa^{(2,1)}$ which becomes
\be
\Gamma_\kappa^{(2,1)}(\omega_1,\omega_2;\vp_1,\vp_2) \rightarrow \sqrt{g_b}\, \vp_1\cdot \vp_2\, f_\kappa^\lambda\left(|\vp_1+\vp_2|\right). 
\label{gam3}
\ee
One can  check that the NLO approximation preserves the KPZ symmetries.

The NLO approximation resembles the simple approximation achieved in Ref.\ \cite{canet10}, to which we refer as leading order (LO). Indeed,  the LO approximation is recovered by setting the entire function $f^\lambda_\kappa(\omega,p)\equiv 1$ (not only at $p=0$) {\it in all dimensions} in the NLO flow equations. 
In dimension one, the two approximations LO and NLO coincide as $f_\kappa^\lambda=1$ according to Eq.\ (\ref{eq:d1flam}) -- unless one chooses to relax this constraint at NLO, as detailed in Appendix C.

\subsection{Flow equations}

We now derive the flow equations for the three running functions $f_\kappa^\xx$, $X\in \{ \lambda, \nu,D \}$. For this, we consider the flow equations of the 2-point vertex functions, given by Eq.\ (\ref{dkgam2}). 
The propagator $G_\kappa(\omega,\vq)$ is defined in Eq.\  (\ref{eq:propag}).  
Using the expressions of the 2-point functions Eqs.\  (\ref{anzgam2}) in  the NLO approximation, {\it i.e.}\ with the replacement  (\ref{eq:nloreplace}), and of the regulator $R_\kappa$ given by Eq.\  (\ref{Rk}), we get
\begin{equation}
G_\kappa(\omega,\vq) =\frac{1}{P_\kappa(\omega,q)}\left(\!\! \begin{array}{cc}
2 k_\kappa(q) &  Y_\kappa(\omega,q)\\
 Y^*_\kappa(\omega,q) & 0 
\end{array}\!\!\right) , \label{propag}
\end{equation}
where
\begin{subequations}
\begin{align}
k_\kappa( q) &= f^\td_\kappa( q)+ D_\kappa \,r( q^2/\kappa^2) , \\
l_\kappa( q) &= q^2 ( f^\nu_\kappa\left( q\right)+  \nu_\kappa\, r( q\,^2/\kappa^2) ), \\
Y_\kappa(\omega,q)&= i \omega \,f^\lambda_\kappa(q)+ l_\kappa( q)  , \\
P_\kappa(\omega, q)& = (\omega \,f^\lambda_\kappa  \left(q\right))^2 +  (l_\kappa \left(q\right))^2 ,
\end{align}
\end{subequations}
and $q = |\vq\,|$.
The derivative of the regulator matrix (\ref{Rk})  is
\begin{equation}
\label{eq:dRk}
\partial_\kappa R_\kappa(\vq) = 
\left(\!\! \begin{array}{cc}
0                                                 &      {q^2 \partial_\kappa S^\nu_\kappa (q)} \\
 {q^2 \partial_\kappa S^\nu_\kappa (q)} &  -2  { \partial_\kappa S^\td_\kappa (q)}
\end{array}\!\!\right), 
\end{equation}
where we have defined
\begin{subequations}
\begin{align}
S_\kappa^\xx (q)  &=  X_\kappa r(y)  \, \, ,\, \, y = q^2/\kappa^2 \, \, ,\, \,  X \in \{ D,\nu \}  ,\\
\kappa \partial_\kappa S^\xx_\kappa (y)  &= - X_\kappa \, (\etax_\kappa r(y) + 2 y \,\partial_{y}  r(y)).
\end{align}
\label{eq:kdkS} 
\end{subequations}
The expressions for the propagator  (\ref{propag}), for the unique non-vanishing vertex  $\Gamma_\kappa^{(2,1)}$ (\ref{gam3}) and for the derivative of the regulator matrix  (\ref{eq:dRk}) are then substituted in the NPRG equation (\ref{dkgam2}) to get the flow equations for the (momentum {\it and} frequency dependent) 2-point functions $\Gamma_\kappa^{(1,1)}(\omega,\vp)$ and $\Gamma_\kappa^{(0,2)}(\omega,\vp)$. The flow equations for the functions $f^\xx_\kappa$ are  deduced following Eq.\  (\ref{anzgam2}). We finally obtain
\begin{widetext}
\begin{subequations}
\begin{align}
\p_\kappa f_\kappa^\td(\varpi,p) & = 2 g_b f_\kappa^\lambda(p)^2\dis \int_{\omega,\vq} \frac{(\vq\,^2 +(\vp\cdot\vq))^2 \,k_\kappa(Q)}{P_\kappa(\omega,q)^2 P_\kappa(\Omega,Q)}\Bigg\{P_\kappa(\omega,q)\, \p_\kappa S_\kappa^\td(q)-2\, \vq\,^2 \,l_\kappa(q)\,k_\kappa(q)\, \p_\kappa S_\kappa^\nu(q) \Bigg\} , \\
 \p_\kappa f_\kappa^\nu(\varpi,p) & =\dis -2\frac{g_b}{p^2}  f_\kappa^\lambda(p)\int_{\omega,\vq}  \frac{\vq\,^2 +(\vp\cdot\vq)}{P_\kappa(\omega,q)^2 P_\kappa(\Omega,Q)}\Bigg\{-\vp\cdot\vq \, f_\kappa^\lambda(Q)\, l_\kappa(Q)\, P_\kappa(\omega,q) \, \p_\kappa S_\kappa^\td(q)  \nonumber \\
&+ \Big[2\,\vp\cdot\vq \, f_\kappa^\lambda(Q)\, l_\kappa(Q) \,l_\kappa(q)\,k_\kappa(q) +(\vp\,^2 + \vp\cdot\vq)\,f_\kappa^\lambda(q)\, k_\kappa(Q)(\omega^2 \,f_\kappa^\lambda(q)^2-l_\kappa(q)^2 ) \Big]\, \vq\,^2\, \p_\kappa S_\kappa^\nu(q)  \Bigg\} , \\
 \p_\kappa f_\kappa^\lambda(\varpi,p) & =\dis 2\frac{g_b}{\varpi}  f_\kappa^\lambda(p)\int_{\omega,\vq}  \frac{\vq\,^2 +(\vp\cdot\vq)}{P_\kappa(\omega,q)^2 P_\kappa(\Omega,Q)}\Bigg\{-\Omega\,\vp\cdot\vq\, f_\kappa^\lambda(Q)^2\,P_\kappa(\omega,q)\, \p_\kappa S_\kappa^\td(q)  \nonumber \\
& + 2  \Big[\Omega\,\vp\cdot\vq \, f_\kappa^\lambda(Q)^2 \,k_\kappa(q)+\omega\,(\vp\,^2 + \vp\cdot\vq)\, f_\kappa^\lambda(q)^2 \,k_\kappa(Q)  \Big]\,\vq\,^2 \, l_\kappa(q)\, \p_\kappa S_\kappa^\nu(q)\Bigg\} .
\end{align}
\label{eqf}
\end{subequations}
\end{widetext}
To shorten the notation, we introduced $Q = |\vp+\vq|$ and $\Omega=\varpi+\omega$.
Notice that even  if  frequencies have been neglected in the $f_\kappa^\xx$ functions {\it within} the integrands of the flow equations (\ref{eqf}), these integrands still have an explicit (internal and external) frequency dependence, which is polynomial both in the numerators and denominators.  This generates a non-trivial frequency dependence for the flowing functions $f_\kappa^\xx$  on the l.h.s.\  of the NLO flow equations  (\ref{eqf}). 
For consistency,  the frequency independent flowing functions $f_\kappa^\xx(p)$ on the r.h.s.\ of Eqs.\ (\ref{eqf}) are obtained from the frequency dependent ones on the l.h.s.  $f_\kappa^\xx(\varpi,p)$ evaluated at zero external frequency, that is  $ f_\kappa^\xx(p) = f_\kappa^\xx(0,p)$.
The integrals over the internal frequency $\omega$ can be simply carried out analytically (see Appendix A).

\subsection{Dimensionless flow equations}

As we intend to analyze fixed point properties, we introduce dimensionless and renormalized quantities (denoted by a hat).  Momentum and frequency are measured in units of the running cutoff $\kappa$, 
 \begin{equation}
 \tp = p/\kappa \quad , \quad
 \tnu = \varpi/(\nu_\kappa\kappa^2), 
  \label{eq:dimlessVar}  
   \end{equation} 
and we define (dimensionless) renormalized functions as
 \begin{equation}  
 \tf_\kappa^\xx(\tnu,\tp)  = f_\kappa^\xx(\varpi,p)/X_\kappa,
  \label{eq:dimlessFunc}  
 \end{equation}   
using again the notation $X$ to designate the three parameters $D$, $\nu$, and $\lambda$, with the conventions
\begin{align}
&X \in \{D,\nu,\lambda\}, \,\, X_\kappa \in  \{D_\kappa,\nu_\kappa,1\},\,\,  \eta^\xx_\kappa \in  \{\etad_\kappa,\etan_\kappa,0\}. 
\label{eq:notation}
\end{align}
$D_\kappa$ and $\nu_\kappa$ are the running coefficients Eq.\ (\ref{eq:dknukdef}), which  yields by definition
\begin{equation}
 \tf_\kappa^\xx(\tnu=0,\tp=0)=1 \quad \quad \hbox{for all $\kappa$}.
\label{eqnormfa}
\end{equation}
We also introduce the dimensionless flow variable 
\begin{equation}
s = \ln(\kappa/\Lambda) \quad , \quad \partial_s=\kappa\partial_\kappa ,
 \label{eq:RGtime}  
\end{equation}
often referred to as RG ``time''. The initial scale $\kappa = \Lambda$ corresponds to $s = 0$ and  the macroscopic scale  $\kappa \rightarrow 0 $ is obtained in the limit $s \rightarrow -\infty$. 

The flow equations for the dimensionless functions are simply deduced from  Eqs.\ (\ref{eq:dimlessVar},\ref{eq:dimlessFunc}) as
\begin{equation}
\partial_s \tf_\kappa^\xx(\tnu,\tp) \equiv \kappa \partial_\kappa \left[\frac{1}{X_\kappa} f_\kappa^\xx \left(\frac{\varpi}{\nu_\kappa\kappa^2}, \frac{p}{\kappa}\right)\right] ,
\label{eq:dimlessgen}
\end{equation}
 with the notation  (\ref{eq:notation}). This yields
 \begin{align} 
\label{eqfa}
 \partial_s \tf_\kappa^\xx(\tnu,\tp) 
  &= \etax_\kappa \tf_\kappa^\xx(\tnu,\tp)+(2-\etan_\kappa) \tnu \;\p_{\tnu} \tf_\kappa^\xx(\tnu,\tp)\nonumber\\
 &+ \tp \;\p_{\tp} \tf_\kappa^\xx(\tnu,\tp) +\tI_\kappa^\xx(\tnu,\tp).
  \end{align}
The first three terms of Eq.\  (\ref{eqfa}) stems from the dimensional part. The nontrivial nonlinear contribution is captured by the dynamical part encoded in the  dimensionless integrals
 \begin{equation}  
\label{eq:flowint}
 \tI_\kappa^\xx(\tnu,\tp) 
  =  \displaystyle \frac{1}{X_\kappa} (\kappa \partial_\kappa f_\kappa^\xx(\varpi,p)). 
  \end{equation}    
The integrals $\tI_\kappa^\xx$  are given by the r.h.s.\ of Eqs.\ (\ref{eqf}) where the substitutions for dimensionless quantities Eqs.\ (\ref{eq:dimlessVar},\ref{eq:dimlessFunc}) are performed.
Consequently, the bare coupling $g_b$ which appears linearly in  the three integrals is changed to
\begin{equation} 
 g_b \to g_b\,\kappa^{d-2}\, D_\kappa/\nu_\kappa^3  \equiv \tg_\kappa,
\label{eq:runninggkt}
\end{equation}
which is dimensionless. 
In fact, when going to the dimensionless renormalized quantities, all the  dependence in the running coefficients  $D_\kappa$ and $\nu_\kappa$ is  absorbed into the  running parameter $\hat g_\kappa$, which is the only remaining independent coupling. 
Its flow equation is simply deduced from  Eqs.\ (\ref{eq:etaflow},\ref{eq:runninggkt}) as
\begin{equation}
\partial_s \tg_\kappa = \tg_\kappa (d-2+3\etan_\kappa-\etad_\kappa),
\label{eqg}
\end{equation} 
that is  $\tg_\kappa$  evolves only according to its dimensional  flow. 

Following the definitions (\ref{eq:dknukdef},\ref{eq:etaflow}), the running anomalous dimensions $\etan_\kappa$ and $\etad_\kappa$ are obtained from the zero momentum and frequency sector. Evaluating Eq.\ (\ref{eqfa})  for $X=\nu$ and $X=D$ at $\tnu=\tp=0$,
yields
 \begin{subequations}  
 \begin{align} 
  &  \etad_\kappa + \tI_\kappa^\td(0,0) = 0, \\
  &  \etan_\kappa + \tI_\kappa^\nu(0,0)  = 0 ,
 \end{align}
   \label{eq:flownorm2}
 \end{subequations}    
 as $\tf^\nu_\kappa(0,0)=\tf^\td_\kappa(0,0)=1$ according to Eq.\  (\ref{eqnormfa}).

The two integrals  $\tI^\xx_\kappa$ in Eqs.\  (\ref{eq:flownorm2}) both have a linear dependence in the two anomalous dimensions according to Eqs.\ (\ref{eq:kdkS},\ref{eqf}), which we render explicit by writing
\begin{subequations}
  \begin{align}
   \tI_\kappa^\td(0,0) &= \tI^{\td\td}_\kappa \etad_\kappa + \tI^{\td\nu}_\kappa \etan_\kappa + \tI^{\td\xn}_\kappa  , \\
   \tI_\kappa^\nu(0,0) &= \tI^{\nu \td}_\kappa \etad_\kappa + \tI^{\nu \nu}_\kappa \etan_\kappa + \tI^{\nu\xn}_\kappa.
  \end{align}
\label{eq:sys}
\end{subequations}
The expressions of the $\tI^{\xx\xx}_\kappa$ terms are given for completeness in Appendix A. Solving Eqs.\ (\ref{eq:flownorm2},\ref{eq:sys}) for the anomalous dimensions then yields:
\begin{subequations}
\begin{align}
\etad_\kappa &= \frac{\tI^{ \td \nu}_\kappa \tI^{ \nu \xn }_\kappa - \tI^{ \td \xn }_\kappa (1 + \tI^{\nu \nu}_\kappa) }{ (1+\tI^{\td\td}_\kappa) (1+\tI^{\nu \nu}_\kappa) - \tI^{\td \nu }_\kappa \tI^{ \nu \td}_\kappa } , \\
\etan_\kappa &= \frac{ \tI^{ \nu \td}_\kappa \tI^{ \td \xn}_\kappa - \tI^{ \nu \xn}_\kappa (1 + \tI^{\td\td}_\kappa) }{ (1+\tI^{\td\td}_\kappa) (1+\tI^{\nu \nu}_\kappa) - \tI^{\td \nu }_\kappa \tI^{ \nu \td}_\kappa } .
\end{align}
\label{eq:etas}
\end{subequations}
The details to carry out the numerical integration of the flow equations (\ref{eqfa},\ref{eqg}) are summarized in Appendix A.

\section{Fixed-point properties}
\label{RES}

\subsection{KPZ phase diagram}

We integrate the flow equations (\ref{eqfa},\ref{eqg}) in various dimensions and for different initial values of the single dimensionless bare coupling $\tg_\Lambda= \Lambda^{d-2}g_b$ to determine the phase diagram of the KPZ equation, already presented in \cite{canet10,canet11a}. Let us summarize the result.
In $d \leq 2$, the flow always tends to a nontrivial strong-coupling (SC) and fully attractive fixed point with $\tg^\text{\tiny SC}_*\neq 0$, and thus the interface always roughens. In dimensions $d > 2$,  two different regimes  can be reached, depending on the initial bare coupling $\tg_\Lambda$.  Below a critical initial value $\tg_\Lambda^c$, the running coupling $\tg_\kappa$ flows to zero and  the Gaussian fixed point is reached. The corresponding interface is smooth and characterized  by the Edwards-Wilkinson exponents $z=2$ and $\chi=(2-d)/2$.  Above  $\tg_\Lambda^c$, the running coupling $\tg_\kappa$ flows to a strong-coupling fixed-point with $\tg_*^\text{\tiny SC} \gg 1$. It describes a rough  interface with  $\chi>0$ and where the exponent relation Eq.\  (\ref{eq:exporel}) is fulfilled. The critical value  $\tg_\Lambda^c$ separates  the  basins of attraction of these two fixed points. Right at $\tg_\Lambda=\tg_\Lambda^c$, 
 the flow leads to an unstable fixed-point with $0<\tg_*^\text{\tiny RT} <\tg_*^\text{\tiny SC} $   which drives the roughening transition (RT).

In the following, we show how physical observables, which characterize the stationary KPZ growth, can be obtained from the fixed-point solution of the NPRG flow equations. We  present 
 our results for critical exponents,  correlation and response functions and the associated universal  scaling functions and amplitude ratios   in dimensions $d = $ 1, 2 and 3.
Since we consider only  the strong-coupling  behavior, we drop the explicit `SC' label  in the following, keeping in mind that all fixed point quantities, denoted by a star, are obtained at the strong coupling fixed point.

\subsection{Critical exponents}
\label{resexpo}

In this section, we discuss the critical exponents obtained within the LO and NLO approximations
 in dimensions $d=1$, 2 and 3.
We first establish  the connection between the anomalous dimensions $\etad_\kappa$ and $\etan_\kappa$ and the roughness and dynamical critical exponents $\chi$  and $z$.
At a fixed point, the running anomalous dimensions attain their constant fixed point values $\etad_\kappa \rightarrow \etad_*$ and $\etan_\kappa \rightarrow \etan_*$. From  Eq.\  (\ref{eq:etaflow}), we deduce that the running coefficients then acquire  power law behaviors
\begin{equation}
  D_\kappa = D_0 \, \kappa^{-\etad_*}  \quad ,\quad
 \nu_\kappa = \nu_0  \, \kappa^{-\etan_*},
\label{eq:dk0nuk0}
\end{equation}
where $D_0$ and $\nu_0$ are  two  nonuniversal constants. 
On the other hand, the physical critical exponents $\chi$  and $z$ are defined as the anomalous scaling of the frequency and of the correlation function as
\begin{align} 
  \omega \sim \kappa^{z}  \quad , \quad
   C(t, \vx) \sim \kappa^{- 2 \chi} .
\label{eq:wkckscale}
\end{align}
From  Eqs.\    (\ref{eq:dimlessVar},\ref{eq:dimlessFunc}) , the dimensions of the frequency and of the correlation function   in the NPRG framework are
\begin{equation}
 [\omega] = \nu_\kappa\kappa^2   \quad , \quad [C(t,\vx)]  = \kappa^{d-2} D_\kappa/\nu_\kappa .
 \label{eq:wkckdim2}
\end{equation} 
Comparing Eqs.\  (\ref{eq:dk0nuk0},\ref{eq:wkckscale},\ref{eq:wkckdim2}) then yields
\be
z=2-\etan_*  \quad , \quad \chi = (2-d+\etad_*-\etan_*)/2. 
\label{eq:expo}
\ee
On the other hand, the flow equation  (\ref{eqg}) of the coupling $\tg_\kappa$  implies that  any non-zero  fixed point  with $\tg_* \neq 0$  satisfies the relation
\begin{equation}
 (d-2+3\etan_*-\etad_*) = 0 . 
\label{eq:statg}
\end{equation} 
Combining Eqs.\ (\ref{eq:statg},\ref{eq:expo}) then enforces, as expected, the exponents identity
\begin{equation}
 z+\chi=2
 \label{eq:exporel}
\end{equation} 
 at any non-gaussian fixed point.

We hence restrict the following  discussion of  critical exponents to the values of $\chi$.
In  dimension $d=1$, the two anomalous dimensions are equal and  the exact value $\chi=1/2$ is recovered. This result holds both at LO and at NLO, since both approximations coincide in $d=1$ once the time reversal symmetry  is satisfied.
In  dimension $d=2$, the NLO approximation yields $\chi\simeq 0.373(1)$, which is in  better  agreement with the numerical results than the $\chi$ value obtained at LO (see table \ref{tab1}).  In dimension $d=3$ however, 
the value of $\chi$ is not significantly improved from LO to NLO
 and the discrepancy with the numerics remains noticeable.
The results for the exponents are summarized in table \ref{tab1}.

\begin{table}[h]
\caption{\label{tab1} Critical roughness exponent $\chi$ from the NPRG approach  at LO and NLO  approximations in integer dimensions $d$. The values  `Lit.' report  results from other works:  the exact result in  $d=1$,  and in higher dimensions, it corresponds to an average (with standard deviation) over values  obtained in  \cite{tang92b,Ala-Nissila93,castellano99,marinari00,Reis04, Ghaisas06, kelling11}.  The error bars for the calculated NPRG values reflect the variations around the PMS values (see text).
}
\begin{ruledtabular}
\begin{tabular}{lccccc}
$d$ & 1 & 2 & 3 & 4 \\ \hline
$\chi$ (LO) & 1/2 & 0.330(8) & 0.173(5) & 0.075(4) \\
$\chi$ (NLO) & 1/2 & 0.373(1) & 0.179(4) & -- \\
$\chi$ (Lit.) & 1/2 & 0.379(15) & 0.300(12) & 0.246(7) 
\end{tabular}
\end{ruledtabular}
\end{table}

\begin{figure}[tp]
\epsfxsize=8.5cm
\vspace{0.5cm}\hspace{-.5cm}\epsfbox{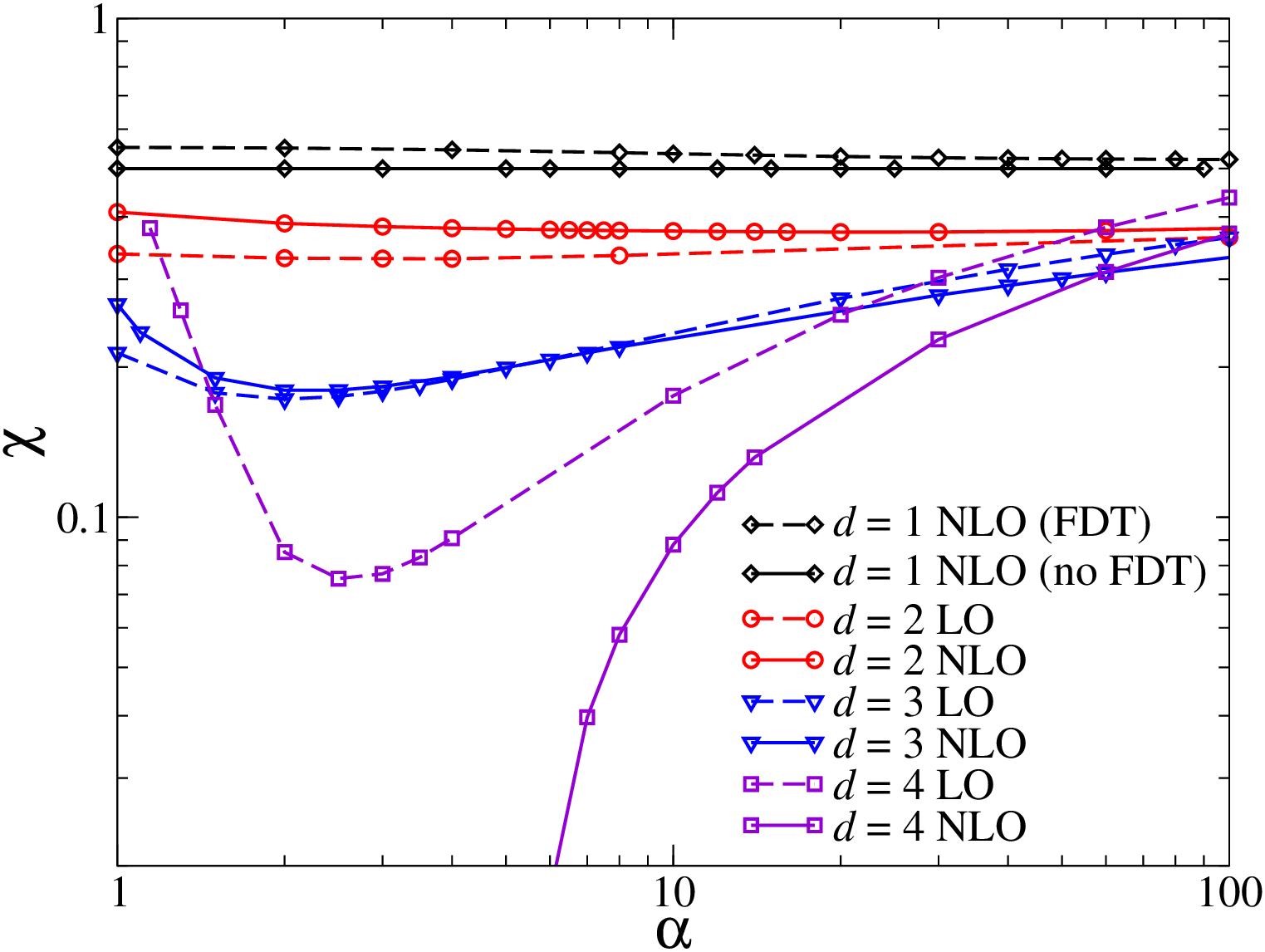}
\caption{(Color online) Variations of the roughness exponent $\chi$ with the cutoff parameter $\alpha$  from the NPRG approach within the LO and NLO  approximations in integer dimensions $d$. In $d=1$ the curves are  computed with the NLO ansatz, either imposing the time reversal Ward identity Eq.\ (\ref{wardfdt}) (FDT) -- in which case  NLO coincides with LO -- or relaxing it (no FDT).
}
\label{fig:expo}
\end{figure}

The  variation of $\chi$ as a function of the cutoff parameter $\alpha$ for different dimensions and approximations are displayed in  Fig.\ \ref{fig:expo} (note the discussion on the role of $\alpha$  in appendix A).  
The values of $\chi$ in table \ref{tab1} represent the minima of the $\chi(\alpha)$ curves of Fig.\ \ref{fig:expo}. The errors reflect the deviations above these values,  when $\alpha$ is varied  within a total range of width 20 ($d=1$), 10 ($d=2$), 1 ($d=3$) and 0.5 ($d=4$) around the minimum values $\alpha_\text{PMS}$.
To estimate the accuracy of the NLO approximation, let us further analyze the $\chi(\alpha)$ curves of Fig.\ \ref{fig:expo}.
Whereas these curves appear essentially flat for dimensions one and two,  their $\alpha$ dependence increases with the dimensionality.
For large  values of $\alpha$ we find that $\chi$ roughly changes logarithmically  with increasing slopes as $d$ grows  both at LO and NLO approximations (except for the $d=1$ NLO with FDT curve, where $\chi=0.5$ for all $\alpha$). Moreover, for $d=$ 2 and 3,  the function $\chi(\alpha)$ exhibits a stable minimum  at both orders NLO and LO.  For $d\gtrsim 3.5$ (curves in non-integer $d$ not displayed),  a PMS keeps existing at LO, whereas it disappears at NLO, which indicates a clear lack of convergence of our ansatz in higher dimensions.  Consequently, the results are expected to be accurate for $d=1$ and $d=2$, seemingly less accurate but still reliable in $d=3$, and not to be trusted for  $d\gtrsim 3.5$.

\subsection{Scaling form of the flowing functions}

We show in this section that the  three functions $\tf_\kappa^\xx$  acquire  a scaling form at the fixed point.
In all dimensions, we observe that all the $\tf_\kappa^\xx$ functions take a nontrivial shape at the strong-coupling fixed point. They are bound to unity by definition (\ref{eqnormfa}) at vanishing 
momentum and frequency and their tails decay algebraically for large $\tnu$ and/or $\tp$.
The fixed point functions $\tf_*^\xx$ are by definition solutions of the stationary equations $\p_s\tf_\kappa^\xx=0$ in Eqs.\  (\ref{eqfa}).
 Besides, we verified that the decoupling property is satisfied.  That is, we determined analytically the limit of the interaction integrals  $\tI_{\kappa}^\xx(\tnu,\tp)$ in the regime  $\tnu \gg 1$ and/or $\tp \gg 1$ and checked that they all  tend to zero. It follows that,  at the fixed point, Eqs.\ (\ref{eqfa}) reduce in the regime $\tnu$ and/or $\tp \gg 1$ to the homogeneous equations
\begin{equation}
 \etax_* \tf^\xx_*(\tnu,\tp)+\tp \;\p_{\tp} \tf^\xx_*(\tnu,\tp)+ (2-\etan_*) \tnu \;\p_{\tnu} \tf^\xx_*(\tnu,\tp) =0,
\label{homog}
\end{equation} 
 with again the conventions Eq.\ (\ref{eq:notation}).
 One can easily show that the  general solutions of Eq.\ (\ref{homog}) take the scaling form
\begin{equation}
 \tf_*^\xx(\tnu,\tp) = \tp^{-\etax_*} \, \baf^\xx \! \left({\tnu}/{\tp^{z}}\right),
\label{hom}
\end{equation}
where $z$ is defined by Eq.\  (\ref{eq:expo}).
The explicit form of the scaling functions $\baf^\xx$ cannot be determined from the homogeneous Eqs.\ (\ref{hom}). However, they can be extracted from the numerical solution of the flow equations (\ref{eqfa}) by
tabulating the values $\tp^{\etax_*}\tf^\xx_*(\tnu,\tp)$ against $\tnu/\tp^{z}$. As shown below, the scaling of the physical correlation and response functions emerges from
the form Eq.\  (\ref{hom}) of the fixed-point solution.

Prior to this, let us determine the asymptotics of the scaling functions.
We denote  $\tau \equiv \tnu/\tp^{z}$ the scaling function argument.
As we find that the fixed point functions $\tf_*^\xx(\tnu,\tp)$ are regular  for all $\tnu$ and $\tp$ values,  constraints for the limits of the scaling functions $\baf^\xx$ can be deduced. 
 At vanishing argument $\tau \to 0$, $\baf^\xx(0)$ has to be finite for the limit $\tnu\to 0$ at fixed $\tp$ to exist and followingly 
\begin{equation}
 \tf_*^\xx(\tnu,\tp) \sim \baf^\xx(0)\tp^{-\etax_*} \quad \text{for} \; \; \tp \gg \tnu^{1/z}.
\end{equation}
At infinite $\tau$, $\baf^\xx$ must behave as
\begin{equation}
 \baf^\xx(\tau)\sim  \baf^\xx_\infty \tau^{-\etax_*/z} \quad \text{for} \; \; \tau\to\infty
\end{equation}
for the limit $\tp\to 0$ at fixed $\tnu$ to exist and with some constant $\baf^\xx_\infty$. Followingly 
\begin{equation}
 \tf^\xx_*(\tnu,\tp) \sim \baf^\xx_\infty\tnu^{-\etax_*/z} \quad \text{for} \; \; \tnu \gg \tp^z.
\end{equation}
Let us recapitulate the explicit leading behaviors of the scaling functions in the limit $\tau\to\infty$ :
\begin{subequations}
\begin{eqnarray}
\baf^\td(\tau) &\sim& \baf^\td_\infty\;\tau^{-\etad_*/z} , \\
\baf^\nu(\tau)&\sim& \baf^\nu_\infty\;\tau^{-\etan_*/z} , \\
\baf^\lambda(\tau) &\sim& \baf^\lambda_\infty = 1 , \label{asympc}
\end{eqnarray}
\label{asymp}
\end{subequations}
where the last constant in Eq.\  (\ref{asympc}) is fixed due to the shift gauged symmetry Eq.\ (\ref{eq:shiftgauge}).

\subsection{Correlation and response functions}
\label{scaling}

The  physical (dimensionful) correlation and response functions 
can be reconstructed from the (dimensionful) 2-point vertex functions (see Appendix B) in the limit $\kappa \to 0$, which  corresponds to the fixed point. 
Let us first express the dimensionful fixed point functions in terms of the dimensionless ones. Following the definitions Eqs.\  (\ref{eq:dimlessFunc}), they are given by
\begin{equation}
f^\xx_*(\varpi,p) = X_\kappa \tf^\xx_*(\tnu,\tp) = X_\kappa \tf^\xx_*\left(\frac{\varpi}{\nu_\kappa \kappa^2},\frac{p}{\kappa}\right).
\end{equation}
The limit $\kappa\to 0$ at fixed $\varpi$ and $p$ is precisely equivalent to the regime $\tnu \gg 1$ and/or  $\tp \gg 1$ where the decoupling occurs, and where $\tf^\xx_*$ scales according to Eq.\  (\ref{hom}). 
Moreover, when the running scale $\kappa$ tends to zero,  the behavior of $D_\kappa$ and $\nu_\kappa$ is controlled by the fixed point, where according to Eq.\  (\ref{eq:dk0nuk0}) they become power laws (with the nonuniversal constants $\nu_0$ and $D_0$).
Hence, the physical dimensionful functions $f^\xx_*$ can be expressed in terms of the fixed point scaling functions $\baf^\xx$  Eqs.\  (\ref{hom}) as
\begin{subequations}
 \begin{eqnarray}
 f^\td_*(\varpi,p) &=& \dis\frac{D_0}{p^{\etad_*}} \baf^\td\left(\varpi /( \nu_0 p^{z})\right) , \\
 f^\nu_*(\varpi,p) &=& \dis\frac{\nu_0}{p^{\etan_*}} \baf^\nu\left(\varpi /( \nu_0 p^{z})\right) , \\
 f^\lambda_*(\varpi,p) &=& \dis\baf^\lambda\left(\varpi /( \nu_0 p^{z})\right) ,
\end{eqnarray} 
\end{subequations}
where the dimensionless argument has been expressed  using Eq.\  (\ref{eq:dimlessVar}) in terms of the dimensionful variables as
\begin{equation}
\tau = \tnu/\tp^{z}  = \varpi /( \nu_0 p^{z}).
\end{equation} 
Finally, according to Eqs.\ (\ref{anzgam2},\ref{eq:expo}), the 2-point vertex functions write
\begin{subequations}
\begin{align}
\Gamma^{(1,1)}_*(\varpi,\vp) &= \dis \nu_0 \,p^z\left( i \tau \baf^\lambda\left(\tau\right) +  \baf^\nu\left(\tau\right) \right) , \\
\Gamma^{(0,2)}_*(\varpi,\vp) &= \dis  - \frac{2 D_0}{p^{d-2+3\chi}}\,\baf^\td\left(\tau\right) .
\end{align}
\end{subequations}

On the other hand, the correlation and response functions  are related to the  2-point vertex functions at the fixed point via (see Appendix B): 
 \begin{subequations}
\begin{eqnarray}
C(\varpi,\vp) &=&   - \frac{\Gamma^{(0,2)}_*(\varpi,\vp)}{|\Gamma^{(1,1)}_*(\varpi,\vp)|^2} \label{Cscal1} , \\
G(\varpi,\vp) &=&  \dis\frac{\Gamma^{(1,1)}_* (\varpi,\vp)}{|\Gamma^{(1,1)}_*(\varpi,\vp)|^2} \label{Gscal1} .
\end{eqnarray}
\label{CGscal1}
\end{subequations}
From these relations we deduce that the  physical correlation and response functions take in the stationary regime the scaling forms
 \begin{subequations}
\begin{eqnarray}
C(\varpi,\vp) &=&  \frac{2}{p^{d+2+\chi}}  \frac{D_0}{\nu_0^2} \Frond(\tau)\label{Cscal} , \\
G(\varpi,\vp)  &=&  \frac{1}{\nu_0\,p^z} \Grond(\tau)\label{Gscal} ,
\end{eqnarray}
\label{eq:CGscal} 
\end{subequations}
where the two scaling functions $\Frond$ and $\Grond$ are defined as 
 \begin{subequations}
\begin{eqnarray}
\Frond(\tau) &=& \frac{\baf^\td(\tau)}{(\tau \,\baf^\lambda(\tau))^2 + \baf^\nu(\tau)^2} , \\
\Grond(\tau) &=& \frac{i\tau \baf^\lambda(\tau) + \baf^\nu(\tau)}{(\tau \,\baf^\lambda(\tau))^2 + \baf^\nu(\tau)^2}  .
\end{eqnarray}
\label{eq:defFG}
\end{subequations}

The asymptotics of the scaling functions   $\Frond$ and $\Grond$ in the limit $\tau\to\infty$ can be simply deduced from the asymptotics of the $\baf^\xx$ functions in Eq.\  (\ref{asymp}) as 
\begin{subequations}
\begin{eqnarray}
\Frond(\tau)&\sim& \baf^\td_\infty \;\tau^{-(d+2+\chi)/z} , \\
\ima(\Grond(\tau))&\sim& \tau^{-1} , \\
\real(\Grond(\tau))&\sim & \baf^\nu_\infty \;\tau^{-(2+\chi/z)} .
\end{eqnarray}
\label{asympCG}
\end{subequations}
At vanishing $\tau$ one obtains
\begin{subequations}
\begin{eqnarray}
\Frond(0)&=& \baf^\td(0) / (\baf^\nu(0))^2 , \\
\ima(\Grond(0))&=& 0 , \\
\real(\Grond(0))&= & 1 / \baf^\nu(0) .
\end{eqnarray}
\end{subequations}

Note that the  expressions (\ref{eq:CGscal}) were derived within the rescaled theory, 
whereas the KPZ equation is usually studied in its original version with three parameters $\lambda$, $D$ and $\nu$.
According to  Eqs.\  (\ref{rescale},\ref{CGscal1}), the relation between the original (l.h.s.) and the rescaled (r.h.s.)  correlation and response functions is
\begin{subequations}
\begin{align}
C(\varpi,\vp,D,\nu,\lambda) & = \frac{D}{\nu^2}C(\varpi/\nu,\vp,1,1,\sqrt{g_b}) \label{eq:crescale}
\\
G(\varpi,\vp,D,\nu,\lambda) & = \frac{1}{\nu}G(\varpi/\nu,\vp,1,1,\sqrt{g_b}).
\end{align}
\label{GCCW}
\end{subequations}
The  correlation and response functions in the stationary regime of the original theory are hence given by
\begin{subequations}
\begin{eqnarray}
C(\varpi,\vp) &=&  \,\frac{D D_0}{\nu^2 \nu_0^2 } \,\frac{2}{p^{d+2+\chi}}\,\Frond\left(\frac{\varpi}{\nu \nu_0 \,p^z}\right) , \label{Cnorm0} \\
G(\varpi,\vp) &=& \frac{1}{\nu \nu_0} \,\frac{1}{p^{z}} \,\Grond\left(\frac{\varpi}{\nu \nu_0 \,p^z} \right) .
\label{Gnorm0}
\end{eqnarray}
\label{GCnorm0}
\end{subequations}
Equations (\ref{GCnorm0})   prove that the physical correlation and response functions endow a scaling form  in the stationary regime, which  evidences  generic scaling. Let us emphasize that we did not {\it assume} the existence of scaling, it naturally arises from the presence of the fixed point and the form of the solution of the flow equations starting from any reasonable microscopic initial condition.
The scaling functions $\Frond$ and $\Grond$ are hence universal with respect to the bare action -- or equivalently to the initial condition at $\kappa=\Lambda$ of the flow equations -- up to the nonuniversal normalizations. 
 Let us stress that this universal property implies that the scaling functions do not depend on a possibly discrete structure at the microscopic scale, such as for instance the lattice type. However, the Fourier transformations implemented in this work  supposes an underlying flat geometry. Accordingly, for other geometries, the scaling functions may  differ.

Finally, the scaling functions (\ref{GCnorm0}) still  depend on the renormalization scheme
through the parameters $\nu_0$ and $D_0$. This dependence can be removed via an appropriate normalization procedure, which will be described in the following. Prior to this, let us determine universal amplitude ratios, which are independent of the choice of normalization.

\subsection{Amplitude ratios}
\label{sec:ar}

The scaling form of the correlation function in real space is usually expressed as  Eq.\ (\ref{eq:corrfunc}) with the scaling function $F(y)$, behaving as 
\begin{equation}
F(y)  = \left\{
\begin{array}{l l} 
F_0 & y \to 0 \\
F_\infty\, y^{2\chi/z} & y \to \infty , 
\end{array}
\right.\label{ratioF}
\end{equation}
where $F_0$ and $F_\infty$ are constants \cite{hwa91}.
One can build a ratio of these amplitudes which is universal, which 
 we  present now.

Let us first express the $F_0$ and $F_\infty$ amplitudes in term of the correlation function in Fourier space  $C(\varpi,\vp)$ that we have calculated.
For this, we consider the truncated correlation function
\begin{equation}
  \Delta C(t,\vec x) = C(t,\vec x) - C(0,0),
\end{equation}
which  has the same asymptotic behaviors as $C(t,\vec x)$, namely 
\begin{equation}
\Delta C(t, \vec x)  = \left\{
\begin{array}{l l} 
F_0\, x^{2\chi} & t \to 0 \\
F_\infty\, t^{{2\chi}/{z}} & x \to 0, 
\end{array}
\right.\label{ratio}
\end{equation}
and which has a well-defined Fourier transform
\begin{equation}
\Delta C(t,\vec x) =  \int_{-\infty}^{\infty} \! \frac{d \varpi}{2\pi} \! \int \! \! \frac{d^d \vp} {(2\pi)^d}\left(e^{-i(\varpi t-\vp\cdot\vx)}-1\right)C(\varpi,\vec p),
\label{eq:c1fourier}
\end{equation}
where $C(\varpi,\vec p)$ is given by Eq.\ (\ref{Cnorm0}).

We first analyze the limit $x \to 0$.
Noting that $C(\varpi,\vec p)$ in  Eq.\ (\ref{Cnorm0}) only depends on $|\vec p|$, we first change to the variables $\tau = \varpi / (\nu \nu_0 p^z )$ and then to the variables  $u = p (\nu \nu_0 t )^{1/z}$ in Eq.\ (\ref{eq:c1fourier}) to obtain
\begin{equation}
\Delta C(t,0) = 2 \,\frac{D D_0 }{\nu \nu_0 } \left(\nu \nu_0 \right)^{2\chi/z} \,  T_d(\chi) \, t^{2\chi/z} ,
\label{eq:Ct0}
\end{equation}
where $T_d(\chi)$ is defined as
\begin{align}
T_d(\chi) &= \int_{0}^{\infty} \frac{d \tau}{\pi} \int\frac{d^d \vec u}{(2\pi)^d}\,\frac{ \cos(\tau u^z) - 1}{u^{2\chi +d}} \Frond(\tau)  .
\label{eq:tddef}
\end{align}
The expression Eq.\  (\ref{eq:tddef}) can be rewritten as 
\begin{equation}
 T_d = I(2\chi/z) \, u_d(\chi) ,
\end{equation}
with the definition
 \begin{equation}
 I(x) = \int_{0}^{\infty} \frac{d \tau}{\pi}\,\tau^{x}\,\Frond(\tau). 
 \label{defIx}
\end{equation}
The remaining integral $u_d(\chi)$ can be performed analytically, yielding
\begin{align}
u_d(\chi) &=  v_d \int_{0}^\infty du \,\frac{\cos(u^z)-1}{u^{1+2\chi}} \nonumber\\
         & = - \frac{v_d}{z} \left\{ \begin{array}{l l} 
                       \dis  \Gamma\left(-2\chi/z\right)\,\cos(2\pi/z)   & \quad \chi \neq {2}/{3},\chi< 1 \\
      {\pi}/{2}   & \quad \chi =  {2}/{3} .
 \end{array}\right.\nonumber\\ \label{defud}
\end{align}
The factor $v_d$ is related to the angular integration volume:  
\begin{equation}
 v_d^{-1}= 2^{d-1}\,\pi^{d/2}\,\Gamma(d/2) .
\end{equation} 
Comparing the relations Eqs.\ (\ref{ratio},\ref{eq:Ct0}), we deduce the $F_\infty$ amplitude
\begin{equation}
 F_\infty = 2\frac{D D_0}{\nu\nu_0} \left(\nu \nu_0  \right)^{2\chi/z} \, u_d(\chi)\,I(2\chi/z) . \label{eq:finf}
\end{equation}

We now consider the  limit $t \rightarrow 0$. From Eqs.\  (\ref{Cnorm0},\ref{eq:c1fourier}), we obtain
\begin{equation}
 \Delta C(0,\vec x) =2\frac{D D_0}{\nu\nu_0} \, I(0)\,K_d(\vec x,\chi) ,
 \label{C0x} 
\end{equation}
with
\begin{equation}
 K_d(\vec x,\chi) =  \int \frac{d^d \vp}{(2\pi)^d}\,\frac{e^{i\vp\cdot\vx}-1}{p^{2\chi+d}} .
\end{equation}
With the substitution $\vp\cdot\vx = v \cos\vartheta$, the $\vec x$ dependence can be factored out
\begin{equation}
 K_d(\vx,\chi) =|\vec x|^{2\chi}\,  w_d(\chi),
\end{equation}
and the remaining integral $w_d(\chi)$ can be performed analytically
\begin{align}
  w_d(\chi)   &= \frac{v_{d-1}}{2\pi} \int_{0}^{\infty} \! \frac{d v}{v^{2\chi+1}} \!\int_{0}^{\pi} \! d \vartheta \sin^{d-2}\vartheta \left(e^{i v \cos\vartheta }-1\right)\nonumber\\
&=
 \frac{\Gamma(-\chi)}{2^{(2\chi+d)} \pi^{d/2} \Gamma(\chi+d/2)} . \label{defwd}
\end{align}
Comparing the relations Eqs.\ (\ref{ratio},\ref{C0x}), we deduce the $F_0$ amplitude
\begin{equation}
F_0 = 2\frac{D D_0}{\nu\nu_0} \,I(0)\,w_d(\chi). \label{eq:f0}
\end{equation}

With the amplitudes (\ref{eq:finf}) and (\ref{eq:f0}),
 one can build a universal ratio, that we define as
\begin{equation}
R = \dis \left|\frac{F_\infty}{(F_0^2)^{1/z}\,\lambda^{2\chi/z}}\right|. \label{defR}
\end{equation}
Indeed, inserting the expressions of the two amplitudes  in this definition, all nonuniversal factors cancel out to give 
\begin{equation}
R =
\dis\left| \frac{u_d(\chi)\,I(2\chi/z)}{(2\hat g_*)^{\chi/z}\,(|w_d(\chi)|\,I(0))^{2/z}}\right|.
\label{calcR}
\end{equation}
This amplitude ratio is computed from our NPRG correlation functions performing numerically the two integrals $I(0)$ and $I(2\chi/z)$ (see appendix A). 

Our results for $R$ are summarized in table \ref{tab:R}. In $d =$ 1, 
 another amplitude ratio $g(0)$ has been considered in the exact calculation of Ref. \cite{praehofer04}, which is related to $R$
 via $g(0) = 2^{1/3} R$ (see appendix C).
The exact result for  $g(0)$ is the Baik-Rain constant, which is  1.150439.... \cite{Baik00,praehofer04}. From our $R$, we obtain in $d=1$ for $g(0)$ the values  1.23(1) at NLO (this work)  and 1.19(1) at SO \cite{canet11a}. 
The NPRG treatment hence tends to overestimate the universal amplitude ratio  in this dimension. The result is less accurate at NLO than at SO, as expected since NLO is a lower order approximation. The accuracy at NLO is of the same order as that of the MC approximation, which predicts $g(0) = $1.1137 \cite{Colaiori01c,praehofer04}. To our knowledge, no predictions for the universal amplitude ratio $R$ in $d \neq$ 1 are available in the literature.

\begin{table}[h]
\caption{\label{tab2} Amplitude ratio $R$  from the NPRG approach at NLO approximation in integer dimensions $d$. Error bars correspond to deviations when varying the cutoff parameter $\alpha$, using the same intervals  as in Sec. \ref{resexpo}. The exact value in $d =$ 1 is obtained from the Baik-Rain constant \cite{Baik00,praehofer04} and Eq.\  (\ref{eq:g0Rrel}).  To our knowledge no (approximate) results for $R$ in $d = $ 2 and 3 are published.
}
\begin{ruledtabular}
\begin{tabular}{lcccc}
  $d$     & 1   &   2   &   3      \\ \hline
 $R$ NLO & 0.977(1) & 0.940(2) & 0.951(4)  \\ 
 $R$ exact &  0.9131  & -- & -- 
\end{tabular}
\end{ruledtabular}
\label{tab:R}
\end{table}

\subsection{Scaling functions in \textit{d} = 1}
\label{1d}

In one dimension, we  compare the  scaling functions obtained from our two NPRG approximations, namely the NLO (this work) and the SO ansaetze \cite{canet11a}, with the exact scaling functions obtained in Ref.  \cite{praehofer04}, essentially in order to assess the quality of the NLO approximation.
 In Ref.  \cite{praehofer04},  three scaling functions $\fy$, $\fhat$ and $\frond$ are  defined, related one to another by the following Fourier transformations 
\begin{subequations}
\begin{eqnarray}
 \fhat(k) &=& \int_0^\infty \frac{d\tau}{\pi} \,\cos(\tau k^{3/2}) \frond(\tau),\\ 
f(y) &=& \int_0^\infty \frac{dk}{\pi}\, \cos(k y) \fhat(k) , 
\end{eqnarray}
\label{eq:defsp}
\end{subequations}
and normalized using a specific criterion (see Appendix C).
We use  lower-case letters $\frond$, $\fhat$ and $\fy$ to refer to the correspondingly normalized functions. In order to compute the numerical Fourier transformations, our function $\frond$ is smoothened by a third order fit Eq.\  (\ref{eq:fit}) - details can be found  in appendix A and Ref. \cite{canet11a}.

\begin{figure}[h]
\epsfxsize=8.5cm
\vspace{0.5cm}\hspace{-.5cm}\epsfbox{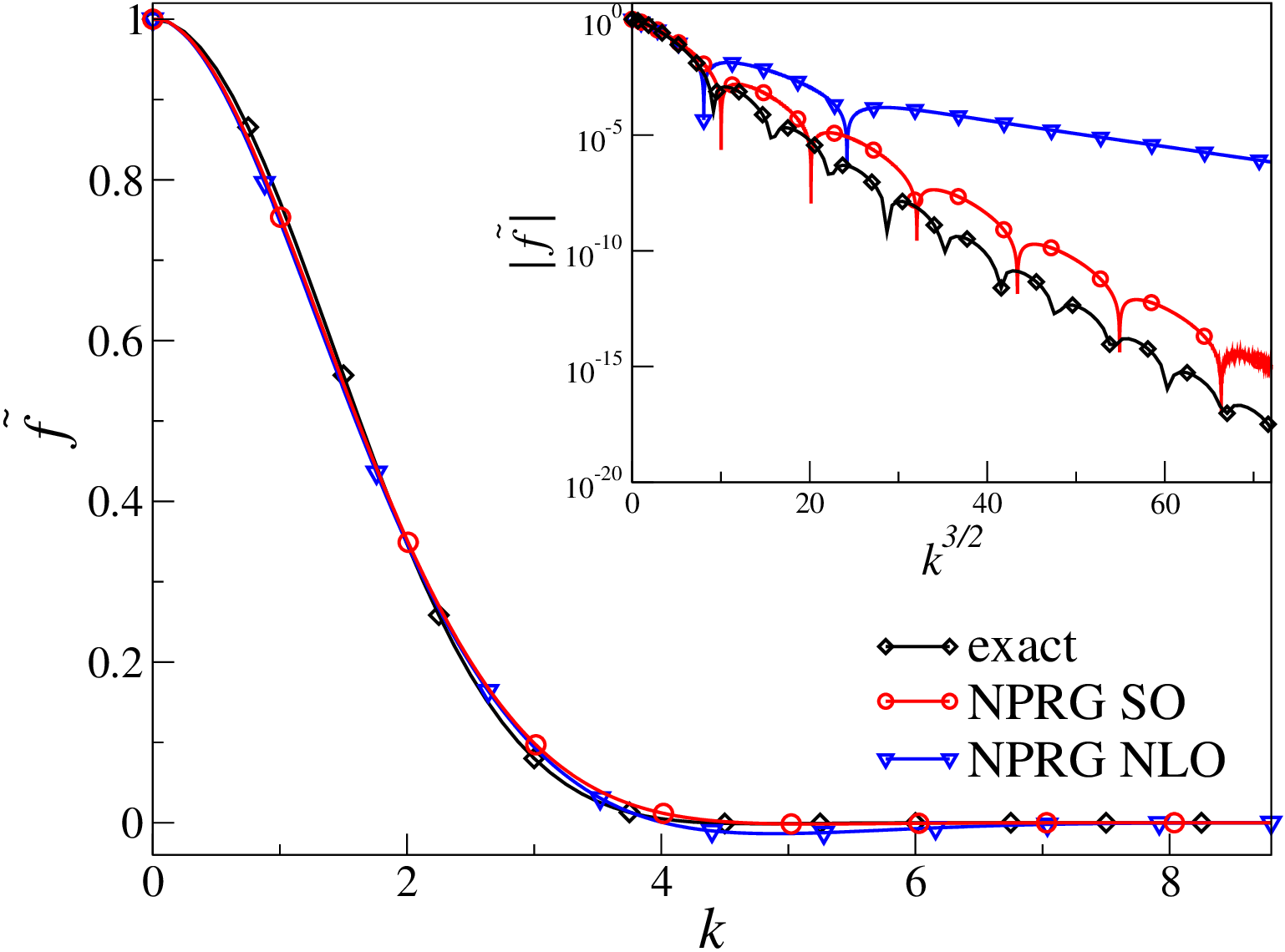}
\caption{(Color online) 
Comparison of the one-dimensional scaling function $\fhat(k)$; exact  result \cite{praehofer04}, NPRG at NLO (respecting FDT, this work) and NPRG at SO  \cite{canet11a}.
The inset shows the stretched exponential behavior of the tail with the superimposed oscillations, developing on the same scale $k^{3/2}$. Note the vertical scale:  this behavior sets in with amplitudes  below typically $10^{-6}$.
}
\label{fig2}
\end{figure}
\begin{figure}[h]
\epsfxsize=8.5cm
\vspace{0.5cm}\hspace{-.5cm}\epsfbox{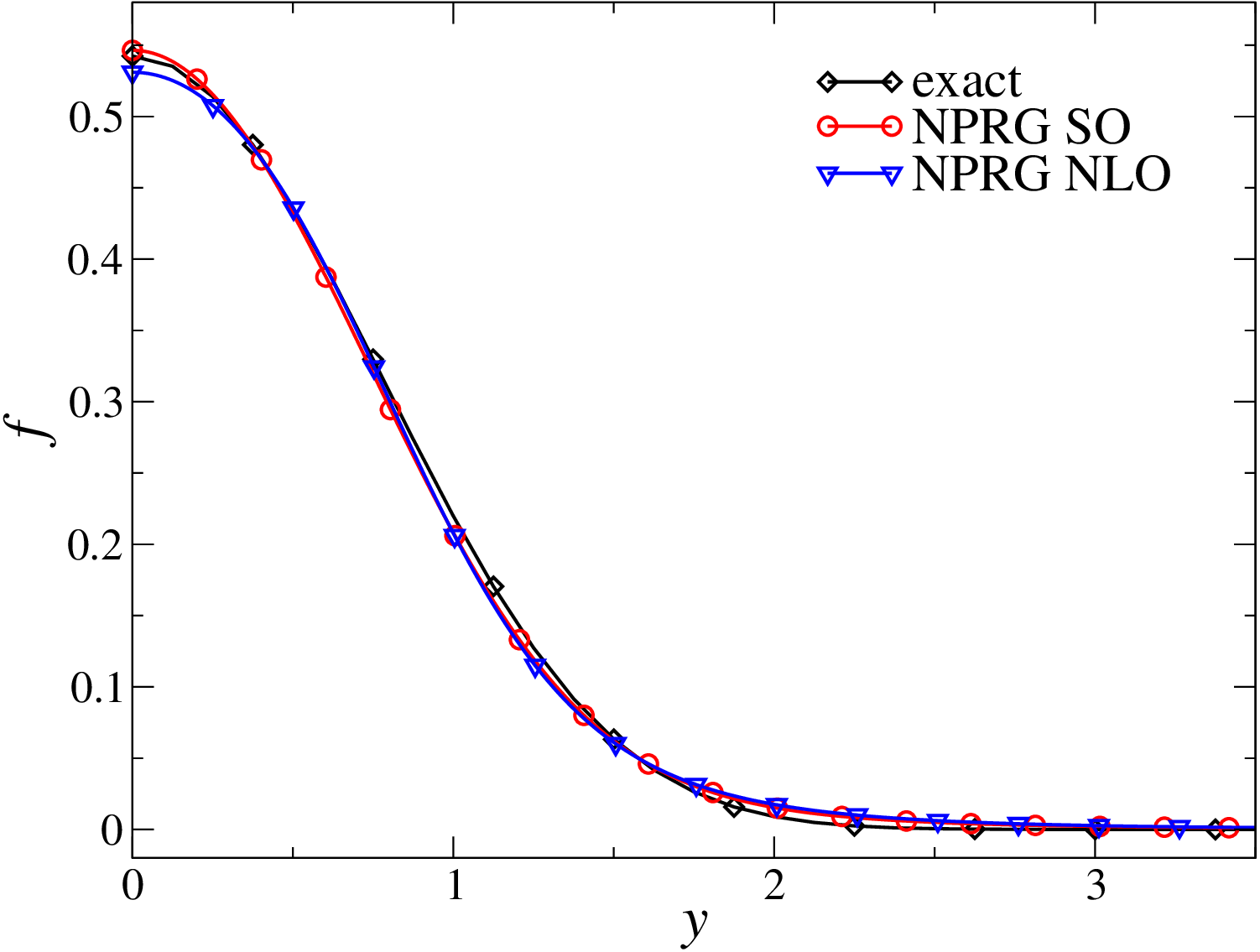}
\caption{(Color online) Comparison of the one-dimensional scaling function $\fy(y)$;  exact  result \cite{praehofer04}, NPRG at NLO (respecting FDT, this work) and NPRG at SO  \cite{canet11a}.}
\label{fig3}
\end{figure}

The scaling functions $\fhat(k)$ and $\fy(y)$ obtained within NPRG are compared with the exact results in Fig.\ \ref{fig2} and \ref{fig3}, respectively. 
One observes that the overall agreement with the exact result is manifestly excellent, both within the  NLO and SO approximations. 
 Examining the details, two slight differences between the NPRG scaling functions can be noticed. First, 
compared to the SO approximation and the exact result, the origin of $\fy$ at NLO is slightly shifted downwards in Fig.\ \ref{fig3}.
Second, the tail of the function $\fhat$ in the NLO approximation is less accurately reproduced than at SO. Indeed,  nontrivial features of this tail are highlighted in \cite{praehofer04}: $\fhat$   decays to zero with a stretched
 exponential tail and with superimposed tiny oscillations around zero, only apparent on a logarithmic scale. A heuristic fit of this behavior for $k\gtrsim 15$ is given by \cite{praehofer04}:
\be
\fhat(k) \sim 10.9 \, k^{-9/4}\sin\left(a_0\,k^{3/2} - 1.937\right)e^{-b_0\,k^{3/2}} .
\label{heuristic}
\ee
Within NPRG, at both NLO and SO approximations,  the decay follows a stretched exponential on the correct scale $k^{3/2}$, but with a less accurate coefficient $b_0$ at NLO. (see table \ref{tab3}).
Moreover, contrarily to the SO approximation, the NLO one essentially misses the oscillations below a typical absolute $\fhat$ value of around $10^{-6}$ (see the inset of Fig.\ \ref{fig2} and \footnote{At NLO, only one oscillation emerges in the tail of $\fhat$ before dying out, see Fig.\ \ref{fig2}. This is related to the fact that the stable complex singularity $z_0$ still exists, as at SO, but it is dominated at NLO by a purely imaginary singularity $z_1=i b_1$ which lies closer to the real axis than $z_0$ (for the fits of order 2 and 3).  It destroys the oscillations at  large $k$, see Appendix A}.)
Additional numbers to further characterize the function $\fhat(k)$ are listed in table \ref{tab3}: position of the first zero $k_0$, coordinates of the negative dip ($k_d$,$\fhat_d$),  oscillation pulsation $a_0$ and decay coefficient $b_0$.

To summarize, the scaling functions in $d$ = 1 obtained from the NLO and the SO ansaetze are very similar, with the SO approximation providing in general a more accurate description, closer to the exact result. This is expected  since the NLO ansatz is obtained from a SO ansatz by truncating the frequency sector. 
 However, the loss of precision remains small at the NLO approximation, given the drastic simplification of the flow equations it entails. We can hence confidently restrict to the NLO approximation in higher dimensions.

\begin{table}
  \caption{Characteristic quantities relative to the one-dimensional scaling function $\fhat$: 
 position of the first zero $k_0$,  coordinates of the negative dip ($k_d$,$\fhat_d$), coefficient of the stretched exponential $b_0$ and pulsation of the oscillations $a_0$ of;  from the exact result \cite{praehofer04}, NPRG at NLO (respecting FDT, this work) and NPRG at SO  \cite{canet11a}.
 The error bars  reflect the  deviations when $\alpha$ is varied within an interval of width 20 around  $\alpha_\text{PMS}$.
}
\begin{ruledtabular}
\begin{tabular}{cccc}
quantity &  exact       &  NLO       &  SO \\\hline
 $k_0$ & 4.36236        & 4.01(5)    & 4.60(6)   \\
 $k_d$ & 4.79079        & 4.89(4)    & 5.14(6)   \\
 $\fhat_d$ & -0.00120   & -0.0147(6) & -0.0018(6)   \\
 $a_0$ &  $\frac{1}{2}$ & --    & 0.28(5)   \\
 $b_0$ &  $\frac{1}{2}$ & 0.134(2)    & 0.49(1)
\end{tabular}
\end{ruledtabular}
\label{tab3}
\end{table}

\subsection{Scaling functions in \textit{d} = 2  and 3}
\label{alld}

The scaling functions $\Frond$ and $\Grond$ are constructed   from our data as defined by Eqs.\ (\ref{eq:defFG}). They are related to the physical correlation and response functions via Eqs.\ (\ref{GCnorm0}). The absolute normalizations of  the scaling functions and their argument  contain nonuniversal factors. In order to define fully universal scaling functions, one must rescale their abscissa and ordinates with an overall factor, fixed by adjusting the
normalizations in an arbitrary conventional way. We here choose a different criterion than that of Ref. \cite{praehofer04} used in the previous section. We introduce three parameters $C_0$, $G_0$ and $\tau_0$ as follows
\begin{subequations}
\begin{eqnarray}
C(\varpi,\vp) &=&  C_0 \,\frac{D D_0}{\nu^2 \nu_0^2 } \,\frac{2}{p^{d+2+\chi}}\,\Frond_N\left(\frac{\varpi}{\nu \nu_0 \,p^z}\,\tau_0\right) ,\label{Cnorm} \\
G(\varpi,\vp) &=& G_0 \frac{1}{\nu \nu_0} \,\frac{1}{p^{z}} \,\Grond_N\left(\frac{\varpi}{\nu \nu_0 \,p^z}\,\tau_0 \right),
\label{Gnorm}
\end{eqnarray}
\label{GCnorm}
\end{subequations}
(the subscript `$N$' denoting normalized functions).
 Comparing Eqs.\ (\ref{GCnorm0},\ref{GCnorm}) yields
\begin{equation}
\Frond_N(\tau) = \frac{1}{C_0} \,\Frond\left(\tau /\tau_0\right) \quad,\quad
\Grond_N(\tau) = \frac{1}{G_0} \,\Grond\left(\tau /\tau_0 \right).
\label{FGnorm}
\end{equation}
We then fix  the three  constants $C_0$, $G_0$ and $\tau_0$ by choosing the following  normalization criteria
\begin{align}
\Frond_N(0) = \Grond_N(0)=1 \quad , \quad
\int_{0}^{\infty} \frac{d \tau}{\pi}\,\Frond_N(\tau) = 1 ,
\label{criterion}
\end{align}
which yields
\begin{align}
C_0 =  \Frond(0), \quad G_0 =  \Grond(0), \quad
\tau_0 = \frac{I(0)}{\Frond(0)}.
\end{align}

\begin{figure}[ht]
\epsfxsize=8.5cm
\vspace{0.5cm}\hspace{-.5cm}\epsfbox{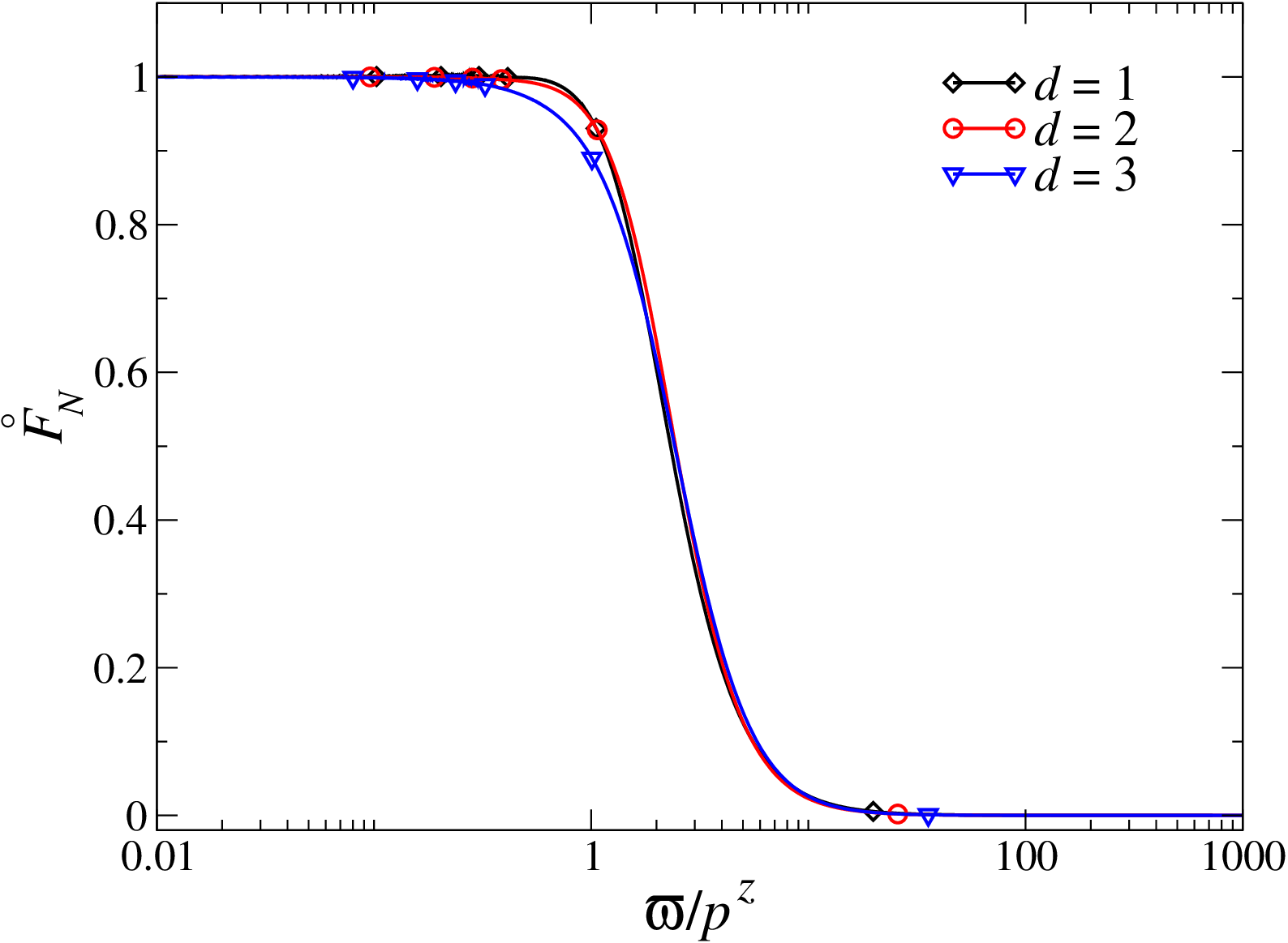}
\caption{(Color online) Normalized scaling function $\Frond_N$ (relative to the correlation function) from NPRG at NLO in dimensions 1, 2 and 3.}
\label{fig4}
\end{figure}

\begin{figure}[ht]
\epsfxsize=8.5cm
\vspace{0.5cm}\hspace{-.5cm}\epsfbox{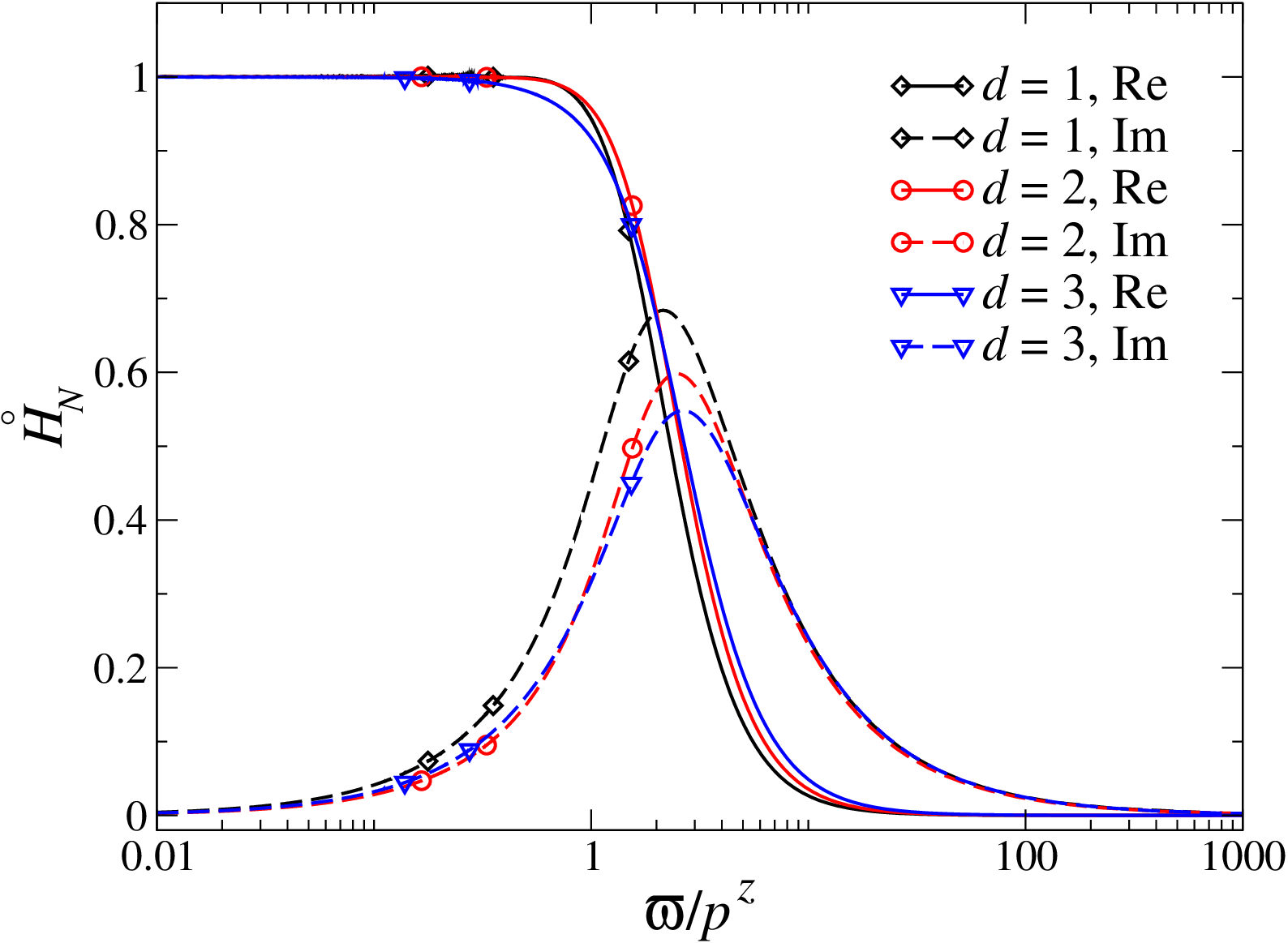}
\caption{(Color online) Real and imaginary parts of the normalized scaling function $\Grond_N$ (relative to the response function) from NPRG at NLO in dimensions 1, 2 and 3.}
\label{fig5}
\end{figure}

The scaling function $\Frond_N$ is displayed in Fig.\ \ref{fig4} and the real and imaginary parts of $\Grond_N$ are displayed in Fig.\ \ref{fig5}. This constitutes the main result of this paper. 
First, let us emphasize that, as  causality is preserved by our approximation scheme and as the regulator term ensures that the functions remain analytic in the complex upper half plane, the real and imaginary parts of the response function are expected to satisfy the Kramers-Kroning relations
\begin{subequations}
\begin{eqnarray}
\real(\Grond(\tau)) &=& \dis\frac{1}{\pi} {\cal P}\,\int_0^{\infty} d\tau'\,\frac{\ima(\Grond(\tau'))}{\tau'-\tau} ,\\
\ima(\Grond(\tau)) &=& -\dis\frac{1}{\pi} {\cal P}\,\int_0^{\infty} d\tau'\,\frac{\real(\Grond(\tau'))}{\tau'-\tau},\label{eq:kk}
\end{eqnarray}
\end{subequations}
where ${\cal P}$ denotes the Cauchy principal value.
 We verified numerically that these relations are perfectly fulfilled.

\begin{figure}[ht]
\epsfxsize=8.5cm
\vspace{0.5cm}\hspace{-.5cm}\epsfbox{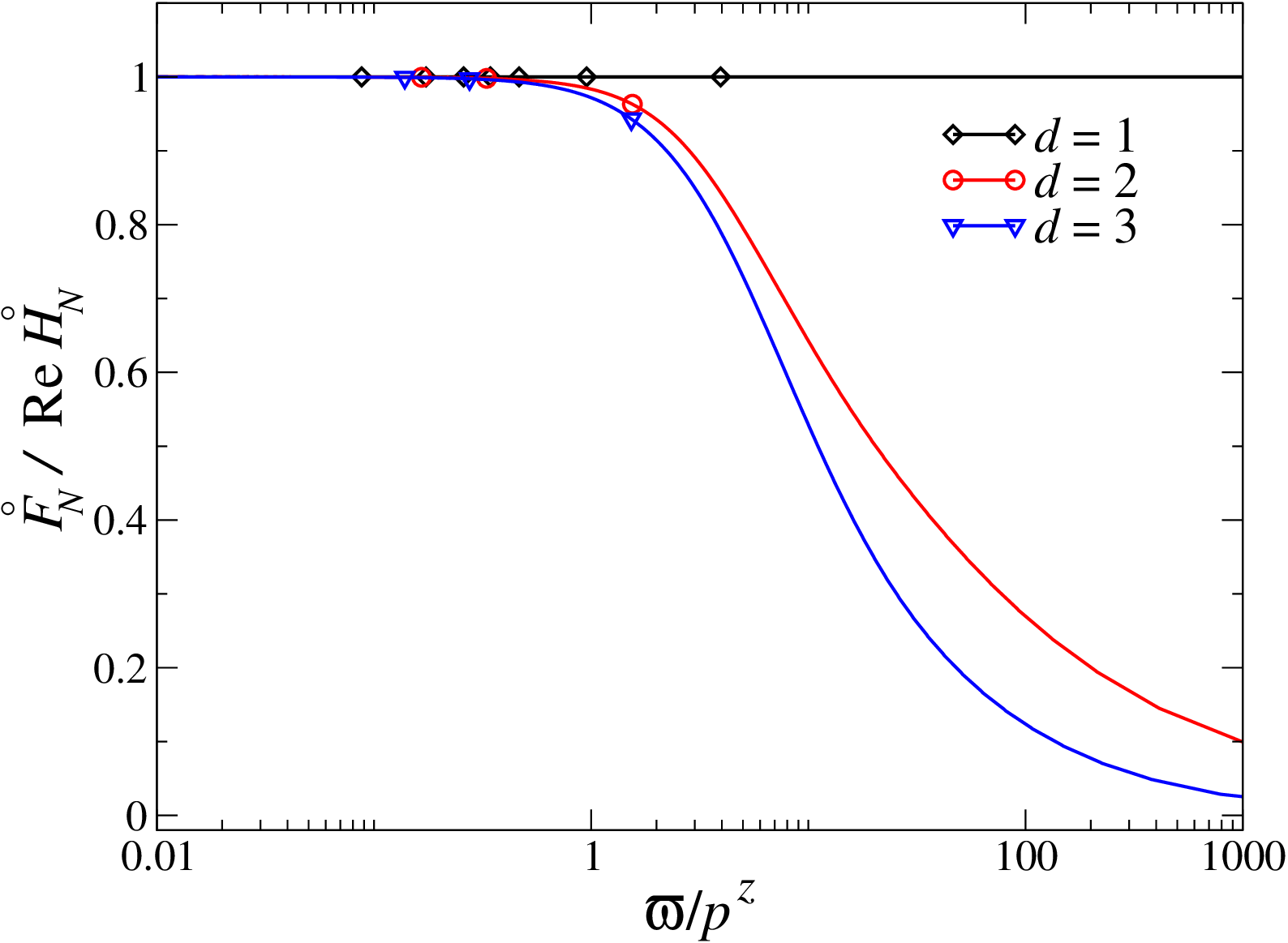}
\caption{(Color online) Ratio of the  scaling function $\Frond_N$ and of the real part of the scaling function $\Grond_N$ in dimensions 1, 2 and 3, which illustrates the departure from the generalized FDT.}
\label{fig6}
\end{figure}

Several quantities  can be studied to further characterize
 these functions. The first one is the universal amplitude ratio $R$, which we already discussed in Sec. \ref{sec:ar}.  The results are reported in table \ref{tab2}.
A second quantity to measure is the departure from the generalized FDT, {\it i.e.}  the ratio $\Frond_N/\real(\Grond_N)$ as a function of $\tau=\varpi/p^z$, which is depicted in Fig.\ \ref{fig6}.
This ratio is constant in $d=1$ where the generalized  FDT holds, that is  $\Frond_N/\real(\Grond_N)(\tau)=1$. In dimensions 2 and 3, the departure from one increases as $\tau$ grows since the powers of the algebraic decay of the two functions $\Frond_N$ and $\real(\Grond_N)$ differ in $d\neq 1$ according to Eqs.\ (\ref{asympCG}).

Lastly, one can examine the structure of the tail of the  function $\Fhat$ defined, in analogy with the one-dimensional case Eq.\ (\ref{eq:defsp}), as 
\be
\Fhat(k) =  \int_0^\infty \frac{d\tau}{\pi} \,\cos(\tau k^{z}) \Frond_N(\tau).
\ee
The asymptotic behavior of $\Fhat$ is determined by the pole structure of $\Frond_N$. To analyze it, we model our data for $\Frond_N$ with the fitting procedure described in Appendix A. We obtain an analogous pole structure as in $d=1$ at NLO. Namely, the singularity lying closer to the real axis is a  pure imaginary one $z_1$. This entails for $\Fhat(k)$ an exponential decay on the scale $k^z$ of the form
\be
\Fhat(k) \sim \exp (-b_1\,k^z) \quad\quad k\to\infty,
\label{eq:tailnlo}
\ee
 and no oscillations (see Appendix A). However, as the absence of oscillations  appeared as an artifact of the NLO approximation in $d=1$, we cannot settle whether the oscillations would persist in higher $d$ in a higher-order approximation \footnote{In fact, in $d=2$ and 3,  the same scenario as the one-dimensional case  occurs.
 the complex singularity $z_0$ with stable coordinates still exists, but is dominated by the pure imaginary pole $z_1$
.}.  The coefficient of the decay is $b_1\simeq 0.61(1)$ in $d=2$ and  $b_1\simeq 0.19(2)$ in $d=3$.

This result can be confronted with the prediction from both the SCE and the MC approximations  \cite{schwartz02,colaiori01a}. For instance, the function  $\hat n(k)$ is studied within the MC approximation \cite{colaiori01a}. It is  related to our function by   $\hat n(k) \propto \Fhat(k^{1/z})$, and it is predicted to decay  exponentially with the asymptotic form
\begin{equation}
\hat n(k) \sim  (B k)^{(d-1)/(2z)}\,e^{-|B k|^{1/z}}\quad k\to \infty
\label{mct}
\end{equation}
with some constant  $B$. It would mean for the function $\Fhat(k) = \hat n(k^z)$ a decay on the scale $k$, and not on the scale $k^z$ as found in our analysis Eq.\ (\ref{eq:tailnlo}). We find the same discrepancy as for the one-dimensional case, where our result agrees with the exact result.

\section{Summary}
\label{conclusion}

In this paper, we analyze the strong-coupling stationary regime of the KPZ equation for an overall flat geometry using NPRG techniques. We work with a simplified version of the full SO (quadratic in the response field) ansatz derived in Ref. \cite{canet11a}. The simplification, that we call NLO, consists in performing an additional approximation in the frequency sector of the flow equations, namely neglecting some frequency dependence  within the integrals of the flow equations, while preserving a nontrivial frequency dependence of the flowing functions. This simplification leads to a drastic reduction of the computing time for the numerical integration of the flow equations, and does not spoil the global quality of the SO approximation, as evidenced in the one-dimensional case.

Indeed, we compute the scaling functions of the one-dimensional growth and confront them with the results obtained with the full SO ansatz \cite{canet11a}  and with the exact results of Ref. \cite{praehofer04}. We show that the NLO approximation essentially preserves the excellent agreement with the exact results that was found at SO, reproducing most of the fine structure of these functions, and with a restricted loss of precision.
The main contribution of this work is the calculation of universal quantities in dimension 2 and 3. We compute  the correlation and the response functions and prove that they take a scaling form in the stationary regime. We provide the associated  scaling functions, and we also calculate universal amplitude ratios. They constitute our most important results. Let us stress  that this is, to our knowledge, the first  predictions for full scaling functions and universal amplitude ratios in dimensions different than one. These predictions could be compared with results from future large scale numerical simulations in $d=2$ and $d=3$. 

 The critical exponents we obtain at NLO compare accurately with numerics in $d=2$ and  reasonably in $d=3$. However, the NLO approximation deteriorates as the dimensionality grows. We provide hints that it becomes unreliable above $d\gtrsim 3.5$, such that the 
  quantitative description of the strong-coupling fixed point properties in dimensions higher than 3 requires a more sophisticated approximation. While high-dimensional correlation and response functions might not be of direct physical relevance, a proper description of these dimensions appears essential in order to probe the existence of an upper critical dimension and to settle on the long-lasting debate on its existence (see {\it e.g}.\ \cite{schwartz12} and references therein). The present ansatz could be improved in several ways. A first option would be to consider the fully frequency-dependent SO \anz (\ref{anznlo}),  which was solved only in $d=1$ so far  \cite{canet11a}. Another option would be to enhance  the response field $\tilde \varphi$ sector, which is truncated at quadratic order in the SO \anz.
It is not {\it a priori}  obvious  which of these two options would be the 
  most relevant in order to improve the high-dimensional behavior of the approximation. This is  left for further investigation.

Finally, only an overall flat geometry was considered in this work. The  influence of other possible geometries in 1+1 and higher dimensions,  in particular on the associated height distribution function,  is definitively an interesting subject to investigate in future work.

\begin{acknowledgments}

The authors  thank the Universidad de la Rep\'ublica and the LPMMC (Grenoble) for funding and hospitality during important stages of this work. 
 The authors acknowledge the support of the cooperation project ECOS Sud France-Uruguay and the support of  PEDECIBA, ANII-FCE-2694. T.\,K. acknowledges financial support by the Alexander von Humboldt foundation.  The numerical parallel codes were run on the clusters FING and PEDECIBA (Montevideo)  and on the cluster CIMENT (Grenoble).

\end{acknowledgments}

\begin{appendix}

\renewcommand{\theequation}{A\arabic{equation}}
\section*{Appendix A: Numerical implementation}
\setcounter{equation}{0}

\subsection*{Solution of the flow equations}

In this appendix we present our   
implementation to numerically solve the coupled flow equations  (\ref{eqfa},\ref{eqg}). 
First, for $d \geq 2$, the $d$-dimensional integrals over the dimensionless internal momentum $\vec\tq$, which have the general form
\be
{\cal I}(\tp) = \dis \int \frac{d^d \vec \tq}{(2\pi)^d} g(|\vec\tq|)F(|\vec \tp +\vec\tq |), 
\label{eq:pInt}
\ee
can be simplified to 2-dimensional integrals.
Using hyperspherical coordinates,  Eq.\  (\ref{eq:pInt}) can be rewritten as
\begin{eqnarray}
{\cal I}(\tp) &=&  \dis\frac{v_{d-1}}{2\pi} \int_{0}^{\infty} d\tq\, \tq^{d-1} \,g(\tq) \nonumber\\
 &\times& \int_0^\pi d\theta\,\sin^{d-2}\theta\, F(\sqrt{\tp^2+\tq^2+2\tp\,\tq\cos\theta}),
 \label{eq:pInt2}
\end{eqnarray}
involving only one radial and one angular integral.

The flowing functions $\tf^\xx_\kappa(\tnu,\tp)$ are discretized on $\tp\times \tnu$ grids in frequency $\tnu$ and momentum $\tp =|\vec \tp|$ space. The grids have a linear spacing $\Delta \tp$ and $\Delta \tnu$ and a maximal range of $\tp_{\text{max}}$ and ${\tnu_{\text{max}}}$. 
The NLO ansatz allows one to perform the frequency integrals analytically (see below).  The momentum integrals are calculated numerically using Simpson's rule.
For $d = 1$, there is a single integral over $\tq$  and for $d \geq 2$ there are two integrals over $\tq$ and $\theta$.
Due to the  presence of the   $\partial_s R_\kappa$ term, the integrands decrease exponentially with $\tq$, which enables one to safely  cut the momentum integral at  a upper finite limit. 
For momenta $\tp + \tq  >\tp_{\text{max}}$  the functions $\tf_\kappa^\xx$ are extended outside the grid using power law extrapolations. This corresponds to the expected asymptotics of the flowing functions, at least close  to the fixed point. 
For momentum coordinates which do not fall onto mesh points, $\tf_\kappa^\xx$ is interpolated using cubic splines. 
The derivative terms $\tp \p_{\tp}$ and $\tnu \p_{\tnu}$ are computed using 5-point differences. 
We studied separately the influence of resolution
($\Delta \tp$, $\Delta  \theta$ and $\Delta {\tnu}$)
and of mesh size ($\tp_{\text{max}}$ and $\tnu_{\text{max}}$) 
on the results.
In the frequency sector, typical grids range up to $\tnu_{\text{max}} = 30$ with a spacing  $\Delta {\tnu} = 1/4$. The precision of the double numerical integral over the momentum is of order $10^{-4}$ for the typical resolution $\Delta \tp=  1/8$ and $\Delta  \theta = \pi / 20$ and an upper integration bound of  $\tp_{\text{max}} / 2 = 10$. 
We checked that the variation of all physical quantities 
by changing the discretization stays of the order $10^{-2}$ smaller than variations coming from the  dependence in the cutoff parameter $\alpha$.

For the propagation in renormalization time $s$, we use explicit Euler time stepping with a typical time step $\Delta s =- 2.5 \times 10^{-5}$.
Starting at $s = 0$ from the bare action ($\tf_\Lambda^\lambda=\tf_\Lambda^\nu=\tf_\Lambda^\td \equiv 1$), the three functions $\tf_\kappa^\xx$ are smoothly deformed 
from their flat initial shapes to acquire their fixed point profiles,  typically after $|s| \gtrsim 10$.
As a convergence criterion we calculate the exponents from Eq.\  (\ref{eq:expo})  and check the deviation from the exponent identity Eq.\  (\ref{eq:exporel}).   
We typically record the fixed point functions at $s=-20$, where this deviation is smaller than $10^{-8}$. We also have to specify the bare initial coupling.
Since the running coupling constant becomes very large in higher dimensions, it is convenient to consider the reduced coupling constant
$\tg_\kappa^{\text{red}} = \tg_\kappa  v_d$, which stays of  order one. A bare initial coupling $\tg_\Lambda^{\text{red}}$ typically ranging between 0.5 and 5 is sufficient to reach the strong coupling fixed point. 
We have checked that the results for universal quantities are independent of the  parameters of the bare action, or equivalently of the initial condition of the flow equations, as expected.

Let us now comment on the analytical integration over the  frequency.
As the function $P_\kappa(\tom,\tq)$ is quadratic in frequency, the integrands in Eqs.\ (\ref{eqf}) are rational fractions in $\tom^2$ with  denominators of order 6 and  numerators of order at most 3. 
The integral over the internal frequency $\tom$ can hence be carried out analytically. For instance
\begin{widetext}
\begin{subequations}
\begin{align}
\int_{-\infty}^{\infty} \frac{ d \hat{\omega}}{2 \pi} \frac{1 }{P_{\kappa}(\hat{\omega},\hat{q}) P_{\kappa}(\hat{\Omega},\hat{Q})} &=
\frac{ \hat{f}^{\lambda}_{\kappa}(\hat{q}) \hat{l}_{\kappa}(\hat{Q}) + \hat{f}^{\lambda} _{\kappa} (\hat{Q}) \hat{l}_{\kappa}(\hat{q}) }{2 \hat{l}_{\kappa}(\hat{q}) \hat{l}_{\kappa}(\hat{Q}) \left( (\hat{f}^{\lambda}_{\kappa}(\hat{q}) \hat{l}_{\kappa}(\hat{Q}) + \hat{f}^{\lambda}_{\kappa}(\hat{Q}) \hat{l}_{\kappa}(\hat{q}))^2 + (\varpi \hat{f}^{\lambda}_{\kappa}(\hat{q}) \hat{f}^{\lambda}_{\kappa}(\hat{Q}) )^2 \right) } \\
\int_{-\infty}^{\infty} \frac{ d \hat{\omega}}{2 \pi} \frac{\hat{\omega}^2 }{P_{\kappa}(\hat{\omega},\hat{q})
P_{\kappa}(\hat{\Omega},\hat{Q})} &=
\frac{ \hat{f}^{\lambda}_{\kappa}(\hat{P}) \hat{l}_{\kappa}(\hat{q}) \hat{l}_{\kappa}(\hat{Q}) + \hat{f}^{\lambda} _{\kappa} (\hat{q}) ((\hat{l}_{\kappa}(\hat{Q}))^2 + (\varpi\hat{f}^{\lambda}_{\kappa} (\hat{q}) ) }{2 \hat{f}^{\lambda} _{\kappa} (\hat{q}) \hat{f}^{\lambda} _{\kappa} (\hat{Q}) \hat{l}_{\kappa}(\hat{Q}) \left( (\hat{f}^{\lambda}_{\kappa}(\hat{q}) \hat{l}_{\kappa}(\hat{Q}) + \hat{f}^{\lambda}_{\kappa}(\hat{Q}) \hat{l}_{\kappa}(\hat{q}))^2 + (\varpi \hat{f}^{\lambda}_{\kappa}(\hat{q}) \hat{f}^{\lambda}_{\kappa}(\hat{Q}) )^2 \right) } \\
\int_{-\infty}^{\infty} \frac{ d \hat{\omega}}{2 \pi} \frac{1 }{(P_{\kappa}(\hat{\omega},\hat{q}) )^2
P_{\kappa}(\hat{\Omega},\hat{Q})} &= 
\frac{ (\hat{f}^{\lambda}_{\kappa}(\hat{q}) \hat{l}_{\kappa}(\hat{Q}) + \hat{f}^{\lambda} _{\kappa} (\hat{Q}) \hat{l}_{\kappa}(\hat{q}) )^2 (\hat{f}^{\lambda}_{\kappa}(\hat{q}) \hat{l}_{\kappa}(\hat{Q}) + 2 \hat{f}^{\lambda} _{\kappa} (\hat{Q}) \hat{l}_{\kappa}(\hat{q}) ) + (\varpi \hat{f}^{\lambda} _{\kappa} (\hat{Q}) \hat{f}^{\lambda} _{\kappa} (\hat{q}) )^2 \hat{l}_{\kappa}(\hat{Q}) \hat{f}^{\lambda} _{\kappa} (\hat{q}) }{4 (\hat{l}_{\kappa}(\hat{q}))^3 \hat{l}_{\kappa}(\hat{Q}) \left( (\hat{f}^{\lambda}_{\kappa}(\hat{q}) \hat{l}_{\kappa}(\hat{Q}) + \hat{f}^{\lambda}_{\kappa}(\hat{Q}) \hat{l}_{\kappa}(\hat{q}))^2 + (\varpi \hat{f}^{\lambda}_{\kappa}(\hat{q}) \hat{f}^{\lambda}_{\kappa}(\hat{Q}) )^2 \right)^2 }
\end{align}
\end{subequations}
\end{widetext}

Finally, we give for completeness the explicit expressions of the integrals at zero external momentum and frequency  involved in the determination of the running anomalous dimensions Eq.\ (\ref{eq:etas}):
\begin{subequations}
\begin{align}
\tI^{\td\td}_\kappa &= - \tg_\kappa \frac{ v_d}{2} \int_{0}^{\infty} \! \! d \tq \, \tq^{d+3} \frac{ r(\tq^2) \, \hat k_\kappa(\tq)}{\tf^\lambda_\kappa(\tq) (\hat l_\kappa(\tq))^3} ,\\
\tI^{\td\nu}_\kappa &= \tg_\kappa \frac{ 3 v_d}{4} \int_{0}^{\infty} \! \! d \tq \, \tq^{d+5} \frac{ r(\tq^2) \,(\hat k_\kappa(\tq))^2}{\tf^\lambda_\kappa(\tq) (\hat l_\kappa(\tq))^4} , \\
\tI^{\td\xn}_\kappa &= \tg_\kappa \frac{ v_d}{2} \int_{0}^{\infty} \! \! d \tq \, \frac{\tq^{d+5} (\partial_{\tq^2} r(\tq^2)) }{\tf^\lambda_\kappa(\tq) (\hat l_\kappa(\tq))^4} \times \nonumber \\
& \qquad \qquad \Bigl[ 3 (\tq \hat k_\kappa(\tq))^2 - 2 \hat k_\kappa(\tq) \hat l_\kappa(\tq) \Bigr] , \\
\tI^{\nu\td}_\kappa &= \tg_\kappa \frac{\pi v_{d+2} }{2} \int_{0}^{\infty} \! \! d \tq \, \frac{\tq^{d+2} r(\tq^2)}{\tf^\lambda_\kappa(\tq) (\hat l_\kappa(\tq))^3} \times \nonumber \\
& \Bigl[ 2 \tf^\lambda_\kappa(\tq) \partial_{\tq} \hat l_\kappa(\tq) - \tq^2 \partial_{\tq} (\tf^\lambda_\kappa(\tq) \hat l_\kappa(\tq) / \tq^2) \Bigr] , \\
\tI^{\nu\nu}_\kappa &= -\frac{\tg_\kappa }{2} \!\! \int_{0}^{\infty} \!\! \! \! d \tq \frac{\tq^{d+3} r(\tq^2)}{\tf^\lambda_\kappa(\tq) (\hat l_\kappa(\tq))^3} \Bigl[ \frac{v_d}{2} \tf^\lambda_\kappa(\tq) \hat k_\kappa(\tq) \! + \! \pi v_{d+2} \times \nonumber \\
& \! \! \! \! \! \Bigl( \tq \tf^\lambda_\kappa(\tq) \partial_{\tq} \hat k_\kappa(\tq) - 2\hat k_\kappa(\tq) (\tf^\lambda_\kappa(\tq) + \tq \partial_{\tq} \tf^\lambda_\kappa(\tq) ) \Bigr) \Bigr] , \\
\tI^{\nu\xn}_\kappa &= \! -\tg_\kappa \!\! \int_{0}^{\infty} \!\! \! d \tq \, \frac{\tq^{d+5} (\partial_{\tq^2} r(\tq^2))}{\tf^\lambda_\kappa(\tq) (l_\kappa(\tq))^3} \!\Bigl[ \frac{v_d}{2} \tf^\lambda_\kappa(\tq) \hat k_\kappa(\tq) \! + \! \pi v_{d+2} \times \nonumber \\
& \!\!\! \! \! \Bigl( \tq \tf^\lambda_\kappa(\tq) \partial_{\tq} \hat k_\kappa(\tq) \! -2 \hat k_\kappa(\tq) (\tf^\lambda_\kappa(\tq) + \tq \partial_{\tq} \tf^\lambda_\kappa(\tq) ) \nonumber \\
& - 2 \tf^\lambda_\kappa(\tq) (\partial_{\tq} \hat l_\kappa(\tq) )/ \tq + \tq \partial_{\tq} (\tf^\lambda_\kappa(\tq) \hat l_\kappa(\tq) / \tq^2)
\Bigr) \Bigr] ,
\end{align}
\end{subequations}
where
\begin{subequations}
\begin{align}
\hat k_\kappa( \hat q) &= \hat f^\td_\kappa( \hat q)+ r( \hat q^2) , \\
\hat l_\kappa( \hat q) &= \hat q^2 ( \hat f^\nu_\kappa(\hat q)+ r( \hat q^2 )).
\end{align}
\end{subequations}

\subsection*{Data analysis}

The physical dimensionful single argument  scaling functions $\Frond(\varpi/p^z)$ and $\Grond(\varpi/p^z)$ (associated with the correlation and response functions) are constructed from the large $\tp$ and/or $\tnu$ sector of the dimensionless two-argument fixed-point functions  $f_*^X(\tnu,\tp)$. In this work,  gridded  $f_*^X$ values between $18 < \tp < 19$ and $28 < \tnu < 29$ where selected to achieve the data collapse.
This  collapse is very accurate in all the dimensions and for all values of $\alpha$ studied.  
This was illustrated in details for the one-dimensional case in Ref.\ \cite{canet11a}. 
However, in order to perform the numerical Fourier transformations Eq.\  (\ref{eq:defsp}) to obtain $\fhat$  and $\fy$, small numerical noise superimposing the raw scaling function $\frond$ causes numerical artifacts. For this reason, we model our scaling functions in terms of a family of fitting functions, as explained in  \cite{canet11a}.
Here we devise a generalization of this Pad\'e-type fit, which especially reproduces the correct asymptotic decay  Eq.\  (\ref{asympCG}). At third order, this fit writes for a general scaling functions $\Frond$:
\begin{equation}  
\Frond_{\text{fit}}(\tau) = \left(\frac{c_{0} + c_{2}\tau^2 + c_{4}\tau^4 + c_{6}\tau^6 }{1 + c_{1}\tau^2 + c_{3}\tau^4 + c_{5}\tau^6 + c_{7}\tau^8 }  \right)^{(d+2+\chi)/2z} ,
\label{eq:fit}
\end{equation}
with the additional constraint 
\begin{equation}  
\baf^\td_\infty = \left(\frac{c_{6}}{c_{7}}  \right)^{(d+2+\chi)/2z} .
\end{equation}
Generalization of this fitting functions to higher orders is straightforward. 
 The family of fitting functions (\ref{eq:fit}), with the appropriate  computed exponents (given in table \ref{tab1}) and $c_{0}=1$ fixed by the normalization criteria (\ref{criterion}), turns out to perfectly adjust to our data for $\Frond_N$ and $\real(\Grond_N)$ in dimensions 2 and 3.  We checked that the third order  fit is sufficient to get a satisfying convergence.
In fact, the same family of fits, with a global exponent one (pure rational fractions),  also perfectly models the function $\ima(\Grond_N(\tau))/\tau$ (which is consistent with  the Kramers-Kroning relations Eq.\ (\ref{eq:kk})). 
 
Moreover, the analytical form (\ref{eq:fit}) allows one to calculate the singularities of this function in the complex plane and determine the asymptotic behavior of  its Fourier transform \cite{canet11a}, defined by   
\begin{equation}
\overline{F}(k) =  \int_0^\infty \frac{d\tau}{\pi} \,\cos(\tau k) \Frond(\tau).
\end{equation}
Indeed, the decay of the tail of $\overline{F}$ is controlled by the singularity of $\Frond$ lying the closest to the real axis.
For instance, in the one-dimensional case at SO, the closest singularity  is  a complex one $z_0 = a_0 + i b_0$ (and symmetrics), which coordinates are robust in successive orders of the fits. This entails for the Fourier transform the exponential asymptotic decay
 $\overline{F}(k) \sim \exp(i z_0) \propto \cos(a_0 k)\exp(-b_0 k)$
and thus for the function $\Fhat$ 
\be
\Fhat(k) = \overline{F}(k^z) \sim \exp(-b_0 k^z)\cos(a_0 k^{z}) \quad k\to\infty.
\ee
If the closest singularity  to the real axis is a pure imaginary one $z_1 = i b_1$ (and symmetrics), there is no oscillation superimposed on the exponential  decay.
 We emphasize that  the exponential decay for $\overline{F}$, and hence the exponential decay on the scale $k^z$ for $\Fhat$
   is rooted in the  $C^\infty$ nature  of the scaling function $\Frond$.
 Only the  existence of a non-analyticity ({\it e.g.} a divergence at the origin)  in $\Frond$ could drive an alternative behavior, which is precluded in our NLO  ansatz.

Let us make a last comment regarding  the numerical calculation of the integrals $I(x)$ defined in Eq.\  (\ref{defIx}), which appear in the  expression of  the amplitude ratio. The  precision can be
 improved by exploiting  the  known asymptotics of $\Frond$. Indeed, as the function $\Frond$ decays as a power law  according to Eq.\  (\ref{asympCG}), the contribution  of the tail can be determined analytically after some sufficiently large crossover argument $\tau_c$, and followingly
\begin{align}
I(x) &= \dis\frac{1}{\pi}\Bigl[\Frond(\tau_c)\,\tau_c^{-(2 \chi+d - z x)/z}\,\frac{z}{2 \chi+d - z x}  \nonumber  \\ 
& \qquad + \int_0^{\tau_c}d\tau \,\tau^x\,\Frond(\tau)  \Bigr],
\label{eq:ixCalc}
\end{align}
where only the bulk integral is computed numerically using a simple trapezoidal rule.

\subsection*{Cutoff dependence}

For the numerical solution of the NPRG flow equations, a  value for the  cutoff parameter $\alpha$ has to be chosen. 
We emphasize that, although the
NPRG flow equation (\ref{dkgam})  is exact and physical quantities are obtained in the $\kappa\to 0$ limit, where the regulator $R_{\kappa=0}$ vanishes, approximations introduce a spurious residual dependence on $R_\kappa$. 
After fixing the functional form of the regulator, this dependence can still be assessed by varying the prefactor $\alpha$  of the regulator and observing the   change in the  physical quantities. This procedure provides us with some accuracy estimate and allows us to check the reliability of the  approximation used. According to the principle of minimal sensitivity (PMS), the different values quoted in this work for physical observables correspond to extremum values  with respect to changes in $\alpha$  \cite{canet03a,*canet03b}.
We further expect that the overall $\alpha$ dependence is weak for the approximation to be accurate.

\renewcommand{\theequation}{B\arabic{equation}}
\section*{Appendix B: Correlation and response functions}
\setcounter{equation}{0}

In this appendix, we establish the relation between correlation and response functions on the one hand and 2-point vertex functions on the other hand. Conventions are taken from Sec.\ \ref{NPRG} and Ref.\ \cite{canet11b}.
The 2-point correlation function matrix ${\cal W}^{(2)}$ is defined as
\begin{align}
{\cal W}^{(2)}({\bf x},{\bf x'}) & =\left(
\displaystyle\begin{array}{ll}
\frac{\delta^2 {\cal W}}{\delta j({\bf x})\delta j({\bf x'})}  & \frac{\delta^2 {\cal W}}{\delta j({\bf x})\delta \tJ({\bf x'})}\\
\frac{\delta^2 {\cal W}}{\delta \tJ({\bf x})\delta j({\bf x'})}& \frac{\delta^2 {\cal W}}{\delta \tJ({\bf x})\delta \tJ({\bf x'})}
\label{eq:w2def}
\end{array}
\right)_{j = \tJ = 0} \nonumber \\ 
& =\left(
\begin{array}{ll}
{\cal W}^{(2,0)}({\bf x},{\bf x'})   &      {\cal W}^{(1,1)}({\bf x},{\bf x'}) \\
 {\cal W}^{(1,1)}({\bf x'},{\bf x})   &      {\cal W}^{(0,2)}({\bf x},{\bf x'}) 
\end{array}
\right) .
\end{align}
In  a uniform and stationary field configuration, the momentum and frequency dependence of its Fourier transform simplifies due to translational invariance in space and time as
\begin{equation}
  {\cal W}^{(2)}({\bf q},{\bf q'}) =  {\cal W}^{(2)}({\bf q}) \, (2 \pi )^{d+1}\delta^{(d+1)}({\bf q}+{\bf q'}), 
  \label{transinv}
\end{equation}
where the matrix ${\cal W}^{(2)}({\bf q})$ writes
\begin{equation}
\label{w2}
{\cal W}^{(2)}({\bf q})=\left( 
\begin{array}{ll}
{\cal W}^{(2,0)}({\bf q})    &      {\cal W}^{(1,1)}({\bf q})\\
 {\cal W}^{(1,1)}({-\bf q})   &      {\cal W}^{(0,2)}({\bf q})
\end{array}
\right)  .
\end{equation}

The matrix ${\cal W}^{(2)}$ is obtained in the NPRG approach as the limit when $\kappa\to 0$,
 {\it i.e.}\ at the fixed point, of the renormalized propagator 
$G_\kappa= {\cal W}^{(2)}_{\kappa}$
\begin{equation}
 {\cal W}^{(2)}  =\lim_{\kappa\to 0} G_\kappa = \lim_{\kappa\to 0} \left[\Gamma^{(2)}_\kappa +R_\kappa\right]^{-1} = \left[\Gamma_*^{(2)}\right]^{-1} .
\end{equation}
As
the 2-point matrix $\Gamma^{(2)}_\kappa$ in Fourier space is also diagonal in momentum, {\it i.e.} has the same form as Eq. \ (\ref{transinv}), it can be simply inverted, which yields
\begin{align}
&  {\cal W}^{(2)} ({\bf q})  =\frac{1}{|\Gamma_*^{(1,1)}({\bf q})|^2}\left(
\displaystyle\begin{array}{ll}
-\Gamma_*^{(0,2)}({\bf q})  & \Gamma_*^{(1,1)}({\bf q}) \\
\Gamma_*^{(1,1)}(-{\bf q})  & -\Gamma_*^{(2,0)}({\bf q}) 
\end{array}
\right) .
\label{W2-gamma}
\end{align}

In real space, the correlation function we consider is defined by Eq.\ (\ref{eq:corrfunc}) and the response function is introduced as
\begin{equation}
G(t,\vx)= \left. \frac{\delta \langle h(t ,\vx )\rangle}{\delta \tilde{j}(0 ,0)}\right|_{j = \tilde{j} = 0} = \langle  h(t ,\vx )  \tilde{h}(0,0) \rangle_c.
\end{equation}
Taking their Fourier transforms  yields Eqs.\  (\ref{Cscal1},\ref{Gscal1}).

\renewcommand{\theequation}{C\arabic{equation}}
\section*{Appendix C: KPZ equation in \textit{D} = 1}
\setcounter{equation}{0}

In this appendix, we discuss the special case of the  one-dimensional KPZ equation, which exhibits the additional time-reversal symmetry. As stressed in Sec.\ \ref{sec:NLOansatz}, the SO ansatz reduces in $d=1$ to only one independent running function $\tf_\kappa$ and one anomalous dimension $\eta_\kappa$ according to Eqs.\ (\ref{eq:d1f}-\ref{eq:d1flam}).

\subsection*{Definition of the scaling functions in $d =$ 1}

First, let us give the definitions and normalizations of the scaling functions calculated exactly in Ref. \cite{praehofer04}, which are denoted with lower case letters  and which serves as our reference.
The real space correlation function considered in \cite{praehofer04} is 
\begin{equation}
c(t,\vx)= \langle [ h(t,\vx) - h(0,0)  - t \langle \partial_t h (t,\vx) \rangle ]^2\rangle = - 2 C(t,\vx), 
\label{eq:fac2corr}
\end{equation}
which differs from the  correlation function $C$ defined in Eq.\  (\ref{eq:corrfunc}) by a factor $(-2)$.
 The scaling function associated with it is defined and normalized as
\begin{equation}
g(y) = \lim_{t\to\infty} \frac{c\left(t,\left(2\lambda^2 N t^2\right)^{1/3}y\right)}{\left(\lambda N^2 t / 2\right)^{2/3}} \label{normg} ,
\end{equation}
where $N= a D/\nu$ is a normalization factor with an arbitrary dimensionless number $a$. 
In fact, the second derivative of this function is mainly considered in Ref.\ \cite{praehofer04}  
\begin{equation}
 \fy(y) = \frac{1}{4} g''(y),
 \label{eq:fgRel} 
\end{equation}
as well as two other scaling functions related by the following Fourier transformations 
\begin{subequations}
\begin{align}
\fhat(k) &= 2\int_0^\infty dy\, \cos(k y) \fy(y), \\
 \frond(\tau)&= 2 \int_0^\infty dk \,\cos(k \tau) \fhat(k^{2/3}),
\end{align} \label{eq:definvsp}
\end{subequations}
which can be inverted according to Eqs.\ (\ref{eq:defsp}).

From Eqs.\ (\ref{normg}) and (\ref{eq:fgRel}) one deduces
\begin{equation}
f(y) = \lim_{t\to\infty} \frac{\left(2 \lambda^2 N t^2 \right)^{1/3}}{2 N} \partial_x^2 \left. \! c(t,x) \right|_{x = \left(2 \lambda^2 N t^2 \right)^{1/3} y} ,
\end{equation}
and using  Eq.\   (\ref{eq:definvsp})  we get
\begin{align}
& \fhat(k) = \lim_{t\to\infty} \frac{- k^{7/2}}{2 N t \left(2\lambda^2 N t^2\right)^{2/3}} \times \nonumber \\
 & \quad \int_0^\infty \frac{d\tau}{\pi} \cos(\tau k^{3/2}) \, c\!\left(\frac{\tau k^{3/2}}{t},\frac{k}{\left(2\lambda^2 N t^2\right)^{1/3}}\right) , 
  \label{eq:fhatSpohn}
\end{align}
where the correlation function on the r.h.s.\ of Eq.\  (\ref{eq:fhatSpohn}) is now the Fourier transform of $c(t,x)$.
Comparing the previous equation with the definition for $\frond(\tau)$ in Eq.\  (\ref{eq:defsp}) finally yields  the relation between the Fourier transformed  correlation function $c(\varpi,p)$ and the scaling function $\frond(\tau)$ as 
\be
 \frond(\tau) = \lim_{t\to\infty}\frac{- k^{7/2}}{2^{5/3}\lambda^{4/3} N^{5/3}t^{7/3}} \, c\left(\tau\frac{k^{3/2}}{t},\frac{k}{\left(2\lambda^2 N t^2\right)^{1/3}}\right) .
 \label{eq:fcSpohn}
\ee

\subsection*{Normalization in $d =$ 1}

We implicitly work in the long time and long distance limit where scaling occurs. Our  scaling function $\Frond$ is related to the  correlation function $c(\varpi,p)$ by
\begin{equation}
  c(\varpi,p) =  (-2) \frac{D}{\nu^2}  \frac{D_0}{\nu_0^2} \frac{2}{p^{7/2}}  \, \Frond\left(  \frac{1}{\nu} \frac{\varpi}{\nu_0 p^{3/2}} \right) ,
   \label{eq:1dcF}
\end{equation}
according to Eq.\ (\ref{Cnorm0}) and with  the additional factor $(-2)$ coming from Eq.\  (\ref{eq:fac2corr}).
Substituting Eq.\  (\ref{eq:1dcF}) into Eq.\  (\ref{eq:fcSpohn}), and recalling that $\nu_0=D_0$ due to Eqs.\ (\ref{eq:d1f},\ref{eq:dk0nuk0}), we obtain 
\begin{equation}
  \frond(\tau) = 2 \left(\frac{ 2 \hat g_* }{a} \right)^{1/2}\, \Frond\left(\left(2 a \hat g_*\right)^{1/2}\tau\right).
  \label{eq:1dprenorm}
\end{equation}
The pure numerical factor $a$ is then fixed according to some normalization criterion, which was chosen in \cite{praehofer04} as
\be
\fhat(0) = 1, \label{normfhat}
\ee
which induces
\begin{equation}
   a = 2 \int_0^\infty \frac{d\tau}{\pi} \, \Frond(\tau) \equiv 2 I(0).
\end{equation}
Inserting $a$ in Eq.\  (\ref{eq:1dprenorm}) then finally yields
\begin{equation}
  \frond(\tau) = 2 \left(\frac{\hat g_*}{I(0)}\right)^{1/2} \Frond \left(2 \left( \hat g_* I(0) \right)^{1/2} \tau \right),
  \label{eq:1dnormfF}
\end{equation}
that we use in the one-dimensional section \ref{1d} to compare our NPRG results with the exact  scaling functions.

\subsection*{Amplitude ratio in $d =$ 1}

In Ref.\ \cite{praehofer04}  the universal amplitude ratio is defined as
\begin{equation}
 g(0) = 4 \dis\int_0^{\infty} dy\,y\,\fy(y) .
 \label{eq:baikrain}
\end{equation}
The exact result for $g(0)$ is the Baik-Rain constant and its numerical value is
$g(0) = 1.15039...$ \cite{Baik00,praehofer04}. This quantity can be expressed in term of $\frond$ as  \cite{canet11a}
\begin{equation}
  g(0) = \dis \frac{2}{\pi^2}\,\Gamma\left(\frac 1 3\right)\,\int_0^{\infty} d\tau\,\tau^{2/3} \frond(\tau) ,
 \label{eq:defg0}
\end{equation}
once the two integrals are carried out analytically.

Let us work out the relation between $g(0)$ and the amplitude ratio  $R$ defined in this work.
The integrals Eqs.\ (\ref{defud},\ref{defwd}) simplify for $d = 1$ and $\chi = 1/2$ to
\begin{subequations}
\begin{align}
u_1(1/2)   &= - \Gamma(1/3)/(2 \pi) ,  \\
 w_1(1/2)  &= - 1/2 ,
\end{align}
\label{eq:uwint}
\end{subequations}
so that  the amplitude ratio $R$ in Eq.\  (\ref{calcR}) reduces to
\begin{equation}
  R =  \frac{\Gamma(1/3)\,I(2/3)}{\pi\,(\hat g_* I(0)^4)^{1/3}} .
   \label{eq:defRF1d}
\end{equation}
On the other hand, combining Eqs.\ (\ref{eq:1dnormfF},\ref{eq:defg0}) yields
\begin{equation}
  g(0) = 2^{1/3} \frac{\Gamma(1/3)\,I(2/3)}{\pi\,(\hat g_* I(0)^4)^{1/3}},
  \label{eq:calcg0}
\end{equation}
from which one  deduces
\begin{equation}
 R = g(0) / 2^{1/3} .
 \label{eq:g0Rrel}
\end{equation}
Note that Eq.\   (\ref{eq:g0Rrel}) can be used to calculate  the amplitude ratio $g(0)$  in $d = 1$ even if $\chi \neq 1/2$, as $R$ is well-defined is this case. This is  relevant in the next section.

\subsection*{Relaxed FDT constraint in $d=1$}

The results obtained at NLO in one dimension are very satisfying and this gives confidence in this approximation.
However, the one-dimensional case is special since the existence of the additional time-reversal symmetry  greatly simplifies the ansatz and the flow equations. Indeed, this symmetry imposes the identities $\tf_\kappa^\nu = \tf_\kappa^\td$ and $\tf_\kappa^\lambda =1$. Let us clarify the origin of this last relation.
In generic dimensions, a generic function $\tf^\lambda(\hat\varpi,\hat p)$ is allowed. In $d=1$,  this function is related  by the time-reversal symmetry  to some  vertices that are neglected in  our ansatz.
As a consequence, our ansatz is only compatible with the time-reversal symmetry and FDT relations if one imposes $\tf^\lambda=1$. Accordingly, one may wonder
 about the importance of this  constraint and what happens if one relaxes it.
 
First, let us  illustrate the role of this constraint. Once the identity  $f_\kappa^\lambda=1$ is imposed, the other Ward identities associated with the time-reversal symmetry are automatically satisfied. Indeed, 
 setting $f_\kappa^\lambda=1$ in the integrands of the dimensionless flow equations (\ref{eqfa}), the two integrals $\tI_\kappa^\td(\tnu,\tp)$  and $\tI_\kappa^\nu(\tnu,\tp)$ in Eq.\   (\ref{eq:flowint}) become equal. Hence, starting from time-reversal symmetric initial conditions  $\tf_\Lambda^\td=\tf_\Lambda^\nu$, both flow equations remain identical and the time-reversal symmetry is preserved along the entire flow. 
On the other hand, if the identity $\tf_\kappa^\lambda=1$ is not imposed,  the dimensionless function $\tf_\kappa^\lambda$ starts to flow, which breaks the equality between $\tI_\kappa^\nu$ and $\tI_\kappa^\td$ and the two functions $\tf_\kappa^\nu$ and $\tf_\kappa^\td$ become different. The FDT relation is then violated.

In order to check the influence of the FDT constraint, we compare results obtained from the NLO ansatz in $d = 1$ with and without imposing it -- that is either with  $\tf_\kappa^\lambda=1$ fixed or with  $\tf_\kappa^\lambda(\tnu,\tp)$ flowing respectively. 
\begin{figure}[ht]
\epsfxsize=8.5cm
\vspace{0.5cm}\hspace{-.5cm}\epsfbox{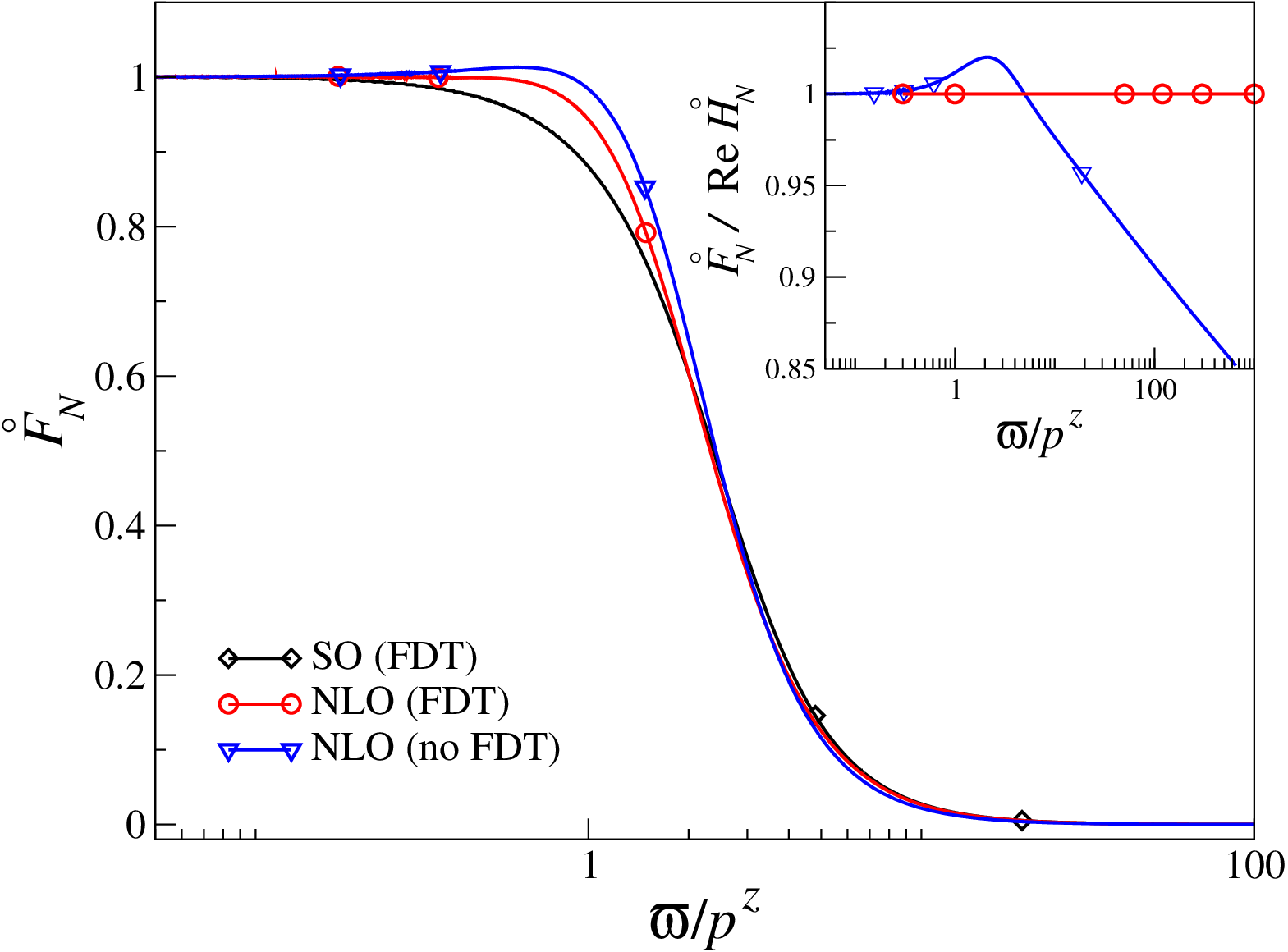}
\caption{(Color online) Comparison of one-dimensional scaling functions $\Frond_N$ obtained from different NPRG approximations: SO, NLO  with FDT and  NLO without FDT.  The scaling function from the SO ansatz \cite{canet11a} also respects FDT. Inset: violation of generalized FDT for the NLO ansatz without the explicit constraint $\tf_\kappa^\lambda=1$.}
\label{fig:noFDT}
\end{figure}
We ran both situations and now discuss the results we obtained. We first observe that the roughness exponent is no longer bound to the exact  result $\chi = 1/2$ at NLO without FDT, but the deviation remains small,  we find $\chi=0.520(5)$. Note that the exponent identity Eq.\  (\ref{eq:exporel}) is still fulfilled as Galilean invariance is preserved. The overall cutoff dependence when varying $\alpha$ is weak as shown in Fig.\ \ref{fig:expo} (no FDT). 
From Eq.\  (\ref{defR}), which is valid in any dimension and for any reasonable values of $\chi$ and $z$, we can compute the universal amplitude ratio $R$. We find $R = 0.871(2) $ (no FDT)  compared to $R= 0.977(1)$ (FDT) and $R =0.9131 $ exact result.
We finally compare in Fig.\ \ref{fig:noFDT} the explicit shapes of the scaling function  $\Frond_N$ with and without FDT, which appear reasonably similar. The inset illustrates the violation of FDT, which increases  for large $\tau$ 
  as the asymptotics of the ratio is given by $\Frond_N/\real(\Grond_N)(\tau)\sim \tau^{-(3+\chi)/(4-\chi)}$ and $\chi\neq 1/2$ without FDT.

\end{appendix}


\end{document}